# Clarifying the relationship between efficiency and resonance for flexible inertial swimmers

## Daniel Floryan[1]† and Clarence W. Rowley[1]

[1]Department of Mechanical and Aerospace Engineering, Princeton University, Princeton, NJ 08544, USA

We study a linear inviscid model of a passively flexible swimmer, calculating its propulsive performance, eigenvalues, and eigenfunctions with an eye towards clarifying the relationship between efficiency and resonance. The frequencies of actuation and stiffness ratios we consider span a large range, while the mass ratio is mostly fixed to a low value representative of swimmers. We present results showing how the trailing edge deflection, thrust coefficient, power coefficient, and efficiency vary in the stiffness-frequency plane. The trailing edge deflection, thrust coefficient, and power coefficient show sharp ridges of resonant behaviour for mid-to-high frequencies and stiffnesses, whereas the efficiency does not show resonant behaviour anywhere. For low frequencies and stiffnesses, the resonant peaks smear together and the efficiency is high. In this region, flutter modes emerge, inducing travelling wave kinematics which make the swimmer more efficient. We also consider the effects of a finite Reynolds number in the form of streamwise drag. The drag adds an offset to the net thrust produced by the swimmer, causing resonant peaks to appear in the efficiency (as observed in experiments in the literature).

**Key words:**

---

## 1. Introduction

A distinguishing feature of nature's swimmers and fliers is the flexibility of their tails and wings, prevailing across a wide range of length scales, time scales, and media. A natural question to ask is whether or not flexibility equips swimmers and fliers with any propulsive advantages over their rigid counterparts, and if so, what characterizes such advantages? The prevailing thinking is that flexibility is indeed a desirable property of a propulsor, but the characterization of its effects, particularly on the efficiency of propulsion, is tenuous.

To be clear, our own interests lie mainly in inertial swimmers characterized by high Reynolds numbers, a large ratio of characteristic fluid mass to body mass, and uniformly distributed passive flexibility. This is in contrast to fliers, for example, where the mass ratio is of order unity and higher, and where the flexibility may be localized. Nevertheless, we will draw upon some of the literature on flight to motivate and guide our analysis.

Passive flexibility has generally been found to lead to thrust and efficiency gains across a range of actuation frequencies, from far below the first natural frequency of the system, to deep into the region of higher order natural frequencies (Alben 2008*b*; Ferreira de Sousa & Allen 2011; Dewey *et al.* 2013; Katz & Weihs 1978, 1979; Quinn *et al.* 2014).

† Email address for correspondence: dfloryan@princeton.edu

In the context of swimming, the efficiency is a measure of how much of the power used to generate the kinematics of a swimmer is converted to useful thrust power. (Although exact definitions vary from one work to the other, they are all in the same spirit.) While thrust generally exhibits local maxima when actuating near natural frequencies (when the system is in resonance), efficiency has been observed to exhibit local maxima below natural frequencies, near natural frequencies, and above natural frequencies (Dewey *et al.* 2013; Moored *et al.* 2014; Quinn *et al.* 2014, 2015; Paraz *et al.* 2016), as well as at frequencies relatively far from a natural frequency (Ramananarivo *et al.* 2011; Kang *et al.* 2011; Vanella *et al.* 2009; Zhu *et al.* 2014; Michelin & Llewellyn Smith 2009). This muddled relationship between efficiency and resonance can be partly explained by an ill-conceived notion of natural frequencies. In some cases (Hua *et al.* 2013; Kang *et al.* 2011; Vanella *et al.* 2009), natural frequencies were based off of an Euler-Bernoulli beam in a vacuum, whereas Michelin & Llewellyn Smith (2009) has shown that the presence of a fluid critically affects the natural frequencies of the system. Some of these studies mistook the added mass of the fluid for drag, leading to an incorrect definition of the total system mass (Vanella *et al.* 2009; Combes & Daniel 2003). In other cases where efficiency and resonance were unrelated, large amplitude motions were considered, leading to a regime of highly nonlinear dynamics where the linear notion of resonance may be inappropriate. We summarize the parameters used in the literature in table 1, translated to correspond with the definitions employed in this work, as defined in section 2. Note that some parameters had to be estimated.

The tacit argument in studies where local maxima in efficiency were observed somewhere near a resonant frequency seems to be that resonance is a condition that improves the efficiency of a system. Although appealing, it is not immediately clear that resonance should unconditionally improve the efficiency of a system. Indeed, most of the works we cite demonstrate local maxima in the input power when the system is actuated at a resonant frequency, not just the thrust, which degrades the efficiency. How resonance affects efficiency is subtle, and should be understood beyond a black-box understanding.

Physical mechanisms unrelated to a fluid-structure resonance have also been offered to explain maxima in efficiency. According to Moored *et al.* (2014), peaks in efficiency occur when the actuation frequency is tuned to a "wake resonant frequency," which is unrelated to any structural frequency. Quinn *et al.* (2015) argued that peaks in efficiency occur when the Strouhal number is high enough that the flow does not separate but low enough that the shed vortices remain tightly packed, the trailing edge amplitude is maximized while flow remains attached along the body, and the effective angle of attack is minimized. In these two works, fluid-structure resonance did coincide with maxima in efficiency. In Ramananarivo *et al.* (2011), peak efficiency was not related to resonance; instead, it was achieved by making use of the nonlinear nature of a drag transverse to the direction of locomotion. The authors argued that efficiency is maximized when the trailing edge is approximately parallel to the total velocity.

In this work, we attempt to clarify the relationship between efficiency and resonance. Resonance is a condition where some property of the system exhibits a maximum; for a passively flexible swimmer, the deflection of its body is one such property. The relation between efficiency and deflection is complicated, making it unclear whether or not resonance of the deflection should result in maximal efficiency. To clarify the matter, we study a passively flexible swimmer using a linear model, valid for small-amplitude motions. Doing so allows us to formally calculate natural frequencies of the coupled fluid-structure system, and to stay in a dynamical regime where the notion of resonance is clear. The linear model cannot account for large-amplitude effects such as separation (especially of the leading edge vortex), but accurately models attached flows (Saffman 1992). Systematically examining the effects of nonlinearity — perhaps separately in the fluid and solid mechanics, and then together — in the context of our linear results would be a useful step forward in delineating exactly what the role of nonlinearity is in swimmers.

| Study | $AR = \frac{w}{L}$ | $Re = \frac{\rho_f U L}{\mu}$ | $R = \frac{\rho_s d}{\rho_f L}$ | $S = \frac{E d^3}{\rho_f U^2 L^3}$ | $f^* = \frac{fL}{U}$ | $h_0$ | $\theta_0$ |
|---|---|---|---|---|---|---|---|
| Alben (2008*b*) | $\infty$ | inviscid | 0 | $10^{-6}$–$10^8$ | $\{1,2,6,25\}/\pi$ | 0 | $\frac{\pi}{720}$ (linear) |
| Dewey et al. (2013) | 2.4 | 7200 | $(1.4$–$34.5) \times 10^{-3}$ | 3.0–78.6 | 0.37–2.99 | 0 | 0.125 |
| Ferreira de Sousa & Allen (2011) | $\infty$ | 851 | — | — | — | 0 | 4°–8° |
| † Hua et al. (2013) | $\infty$ | 200–1800 | 0.5–4 | 2.3–1951 | 0.05–0.5 | 0.25–2 | 0 |
| Kang et al. (2011) | $\infty$ | 9000 | $(4.4$–$33) \times 10^{-3}$ | 3.6, 12.5, 1548 | 0.45–1.55 | 0.388 | 0 |
| Katz & Weihs (1978) | $\infty$ | inviscid | 0 | — | 0.025–0.075 | 6 | 5° |
| Katz & Weihs (1979) | $\ll 1$ | inviscid | 0 | — | 0.025–0.05 | 6 | 5° |
| Michelin & Llewellyn Smith (2009) | $\infty$ | inviscid | 0.2 | $10^{-2}$–$10^4$ | $\{1, 2.5\}/\pi$ | 0.2, 0.4, 1 | 0 |
| Moored et al. (2014) | 2.4 | 7200 | $(1.4$–$34.5) \times 10^{-3}$ | 3.0–78.6 | 0.37–2.99 | 0 | 0.125 |
| ‡ Paraz et al. (2016) | 1 | 0 | low | $\infty$ | 0–17.1 | 0.07–0.24 | 0 |
| Quinn et al. (2014) | 0.77 | 7800–46800 | low | 0.01–1987 | 0–17.1 | 0.1 | 0 |
| Quinn et al. (2015) | 0.77 | 5000–70000 | low | 0.005–1.09 | 0.26–21.74 | 0.06–0.18 | 0°–30° |
| † Ramananarivo et al. (2011) | 1.57 | 88.5–6375 | 1.2–13 | 6.4–44220 | 0.3–26.4 | 0.25, 0.4 | 0 |
| ‡ Vanella et al. (2009) | $\infty$ | 0 | — | $\infty$ | $\infty$ | 2.8 | $\pi/4$ |
| † Zhu et al. (2014) | $\infty$ | 46–454 | 0.1–2 | 6.3–618000 | 0.1–0.85 | 0.4–1.6 | 0 |
| Present study | $\infty$ | inviscid | 0.01 | $10^{-2}$–$10^2$ | $10^{-1}$–$10^2$ | 2 (linear) | 1 (linear) |

TABLE 1. Summary of parameters used in the literature. † denotes studies where the swimmer swam freely (in which case *Re*, *S*, and $f^*$ are dependent variables), and ‡ denotes studies where the freestream velocity was zero. In some cases, parameter values were estimated, and in other cases, parameter values could not be estimated from the information provided (marked as —).

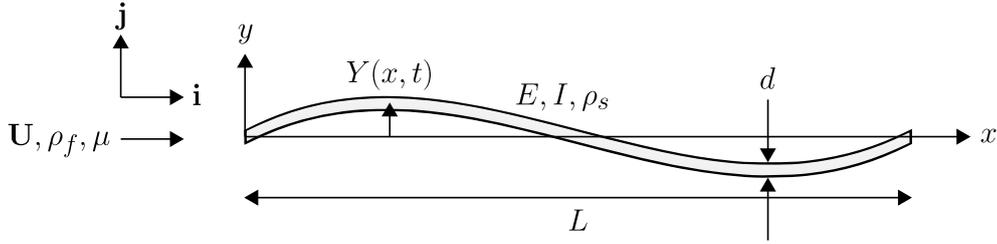

FIGURE 1. Schematic of the problem.

## 2. Problem description

Here, the setup and assumptions are the same as in Moore (2017). Consider a two-dimensional, inextensible elastic plate of length $L$ and thickness $d$. The plate is thin ($d \ll L$), and is transversely deflected a small amount $Y$ from its neutral position, with its slope $Y_x \ll 1$. Under these assumptions, the dynamics of the plate is governed by Euler-Bernoulli beam theory. The plate has uniformly distributed density $\rho_s$ and flexural rigidity $B = EI$, where $E$ is the Young's modulus, $I = wd^3/12$ is the second moment of area of the plate, and $w$ is the width of the plate. The plate is immersed in an incompressible, inviscid Newtonian fluid of density $\rho_f$. There is no flow along the width of the plate, and far from the plate the flow is unidirectional and constant: $\mathbf{U} = U\mathbf{i}$. The setup is altogether illustrated in figure 1.

The motion of the plate alters the velocity field of the fluid, whose forces in turn modify the motion of the plate. The transverse position of the plate satisfies the Euler-Bernoulli beam equation

$$\rho_s dw Y_{tt} + B Y_{xxxx} = w \Delta p, \quad (2.1)$$

where $\Delta p$ is the pressure difference across the plate due to the fluid flow, subscript $t$ denotes differentiation with respect to time, and subscript $x$ denotes differentiation with respect to streamwise position. The fluid motion satisfies the linearized incompressible Euler equations

$$\left.\begin{aligned} \nabla \cdot \mathbf{u} &= 0, \\ \rho_f(\mathbf{u}_t + U\mathbf{u}_x) &= -\nabla p, \end{aligned}\right\} \quad (2.2)$$

where $\mathbf{u} = u\mathbf{i} + v\mathbf{j}$. The above linearization is valid for when the perturbation velocity $\mathbf{u}$ is much smaller than $U$. Since the perturbation velocity depends on the plate's vertical velocity, its slope, and the rate of change of its slope, the linear assumption holds for small-amplitude motions of the plate.

We nondimensionalize the above equations using $L/2$ as the length scale, $U$ as the velocity scale, and $L/(2U)$ as the time scale, yielding

$$\left.\begin{aligned} 2R Y_{tt} + \frac{2}{3} S Y_{xxxx} &= \Delta p, \\ \nabla \cdot \mathbf{u} &= 0, \\ \mathbf{u}_t + \mathbf{u}_x &= \nabla \phi, \end{aligned}\right\} \quad (2.3)$$

{

where
$$R = \frac{\rho_s d}{\rho_f L}, \quad S = \frac{E d^3}{\rho_f U^2 L^3}, \quad \phi = p_\infty - p. \tag{2.4}$$

In the above, $x$, $t$, $Y$, $\mathbf{u}$, and $p$ are now dimensionless, with the pressure nondimensionalized by $\rho_f U^2$. The coordinates are aligned such that $x = -1$ corresponds to the leading edge and $x = 1$ corresponds to the trailing edge. $R$ is a ratio of solid-to-fluid mass, and $S$ is a ratio of bending-to-fluid forces. Note that $\Delta \phi = -\Delta p$.

The fluid additionally satisfies the no-penetration and Kutta conditions, which can be stated as
$$\left. \begin{array}{l} v|_{x \in [-1,1], y=0} = Y_t + Y_x, \\ |v|\,|_{(x,y)=(1,0)} < \infty. \end{array} \right\} \tag{2.5}$$

We impose heaving and pitching motions $h$ and $\theta$, respectively, on the leading edge of the plate, while the trailing edge is free, resulting in boundary conditions
$$Y(-1, t) = h(t), \quad Y_x(-1, t) = \theta(t), \quad Y_{xx}(1, t) = 0, \quad Y_{xxx}(1, t) = 0. \tag{2.6}$$

The fluid motion resulting from the actuation of the leading edge of the plate imparts a net horizontal force onto the plate. In other words, energy input into the system by the actuation of the leading edge is used to generate a propulsive force. The net horizontal force (thrust) on the plate is
$$C_T = \int_{-1}^{1} \Delta p Y_x \, \mathrm{d}x + C_{TS}, \tag{2.7}$$

where $C_{TS}$ is the leading edge suction force (formula given in Moore 2017), and the power input is
$$C_P = -\int_{-1}^{1} \Delta p Y_t \, \mathrm{d}x. \tag{2.8}$$

The leading edge suction force used in Moore (2017) is the limit of the suction force on a leading edge of small but finite radius of curvature, in the limit that the radius tends to zero. The leading edge suction force is a reasonable model of the actual flow when it is attached (Saffman 1992), so we have chosen to include it. In terms of dimensional variables, $C_T = T/(\frac{1}{2}\rho_f U^2 L w)$ and $C_P = P/(\frac{1}{2}\rho_f U^3 L w)$, where $T$ and $P$ are the dimensional net thrust and power input, respectively. Finally, the Froude efficiency is defined as
$$\eta = \frac{\overline{TU}}{\overline{P}} = \frac{\overline{C_T}}{\overline{C_P}}, \tag{2.9}$$

where the overbar denotes a time-averaged quantity.

In this work, we restrict ourselves to actuation at the leading edge that is sinusoidal in time, that is,
$$\left. \begin{array}{l} h(t) = h_0 \mathrm{e}^{\mathrm{j}\sigma t}, \\ \theta(t) = \theta_0 \mathrm{e}^{\mathrm{j}\sigma t}, \end{array} \right\} \tag{2.10}$$

where $\sigma = \pi L f / U$ is the dimensionless angular frequency, $f$ is the dimensional frequency in Hz, $\mathrm{j} = \sqrt{-1}$, and the real part in $\mathrm{j}$ should be taken when evaluating the deflection. Since the system is linear in $Y$, the resulting deflection of the plate and fluid flow will also be sinusoidal in time. We leave the details of the method of solution to Appendix A, noting that all calculations in this work used 64 collocation points. The method to calculate the eigenvalues of the system is detailed in Appendix B, and some useful formulas for the numerical method used are given in Appendix C.

| $Re$ | $R = \frac{\rho_s d}{\rho_f L}$ | $S = \frac{E d^3}{\rho_f U^2 L^3}$ | $f^* = \frac{fL}{U}$ | $h_0$ | $\theta_0$ |
|---|---|---|---|---|---|
| inviscid | 0.01 | $10^{-2}$–$10^2$ | $10^{-1}$–$10^2$ | 2 (linear) | 1 (linear) |

TABLE 2. Parameter values used in this work.

## 3. A note on parameters

It is important to acknowledge that the system parameters we use will critically affect the phenomena we observe. The dissensus in the literature on the relationship between efficiency and resonance may be partly attributed to results being overextended from one dynamical regime to another. We thus take the opportunity here to explicitly state the parameters we employ in this work, as well as to show some resulting qualitative features.

To be clear, the system is parameterized by its Reynolds number $Re$, mass, stiffness, and frequency and amplitude of actuation. Our flow is inviscid, but we will consider some of the effects of a finite Reynolds number later on. As revealed by the nondimensional quantities in (2.4), the mass and stiffness of the system depend on both the solid *and* the fluid. For underwater swimmers, the mass ratio is generally quite low since swimmers are neutrally buoyant but thin; this is in contrast to fliers, for example, where the mass ratio is of order unity and higher. Since our interests lie in swimming flows, we take the mass ratio to be $R = 0.01$ throughout. We vary the stiffness of the system from very flexible ($S \ll 1$) to very stiff ($S \gg 1$), characterized by the stiffness ratio $S$. We vary the frequency of actuation so that it covers multiple natural frequencies of the system. Our system is linear, so scaling the amplitude by some factor will simply scale the flow and deflection fields by the same factor. In this sense, amplitude does not matter in our problem, so we set the heaving and pitching amplitudes so that the maximum deflection of the trailing edge of a rigid plate is equal to the length of the plate. The amplitude affects both thrust and power quadratically, and does not affect efficiency in this linear setting. We do not consider nonlinear effects caused by large amplitudes. The parameters we use in the proceeding sections are summarized in table 2.

As a final note, we point out the effect of the mass of the system. Although we fix the mass ratio to be $R = 0.01$ in the proceeding sections in this work, we take the opportunity here to vary $R$ in order to show how swimmers and fliers may differ, at least qualitatively. In figure 2, we show the efficiency as a function of mass and stiffness ratios for plates heaving and pitching at a reduced frequency $f^* = 1$ (the results are similar to those in figure 11 of Moore (2017), but for slightly different parameter values). The white areas demarcate where the plate produces a net drag (and hence negative efficiency). The relationship between efficiency and stiffness is qualitatively different for low and high mass ratios. At high mass ratios (where the plate is much more massive than a characteristic mass of fluid), the plate does not produce thrust unless the stiffness ratio is high. At $O(1)$ mass ratios, efficiency increases monotonically as the plate becomes stiffer. At low mass ratios, efficiency does not change monotonically with stiffness.

In figure 3, we show the first four natural frequencies of the coupled fluid-structure system as a function of the mass ratio for the limit of large bending velocity compared to flow velocity; we will refer to such frequencies (and, more generally, eigenvalues and eigenfunctions) as quiescent natural frequencies, and refer the reader to Appendix B.2 for more details. To be clear, when we write "natural frequencies" we mean the imaginary parts of the eigenvalues of the system, calculated as in Appendix B. For just the results in this plot, we have changed the time scale such that a flat line indicates that only the mass of the plate (but not the fluid) matters. The nondimensional angular frequency here is $\omega^* = \omega\sqrt{3\rho_s L^4/(4Ed^2)}$, where $\omega = 2\pi f$ is the dimensional angular frequency. For high values of the mass ratio, the natural frequencies scale with the mass of the plate ($\omega^* \sim R^0$). For low values of the mass ratio, however, the natural frequencies scale with the mass of the surrounding fluid ($\omega^* \sim R^{1/2}$). There is also a region where both the characteristic plate and fluid masses must be considered. We also note that for a nonzero incoming flow, the natural frequencies may change (we will show this later).

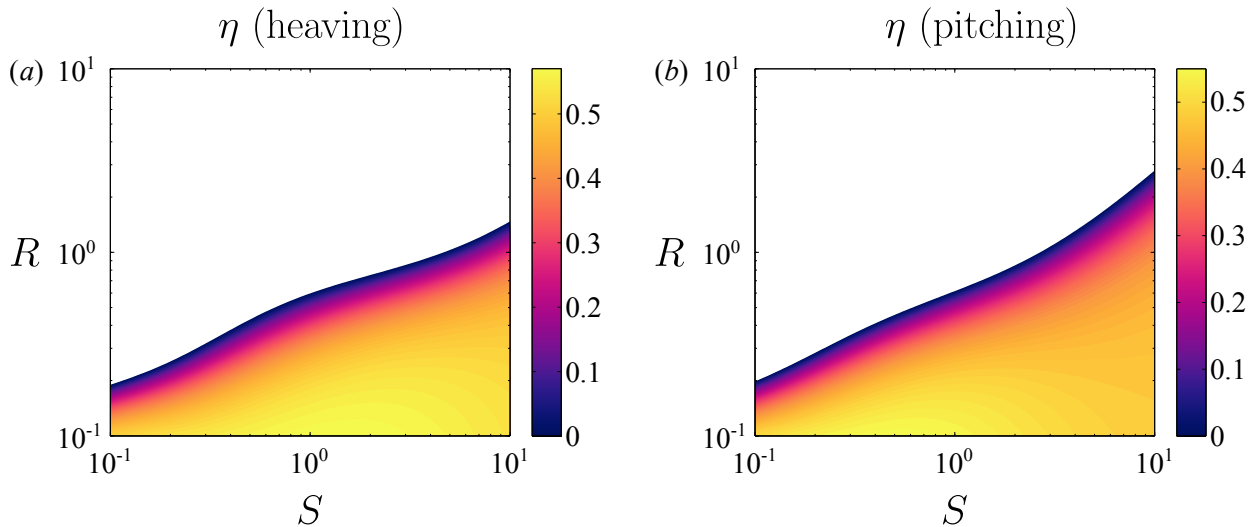

FIGURE 2. Efficiency as a function of mass ratio $R$ and stiffness ratio $S$ for a (a) heaving and (b) pitching plate at $f^* = 1$. Areas with negative efficiency have been whited out.

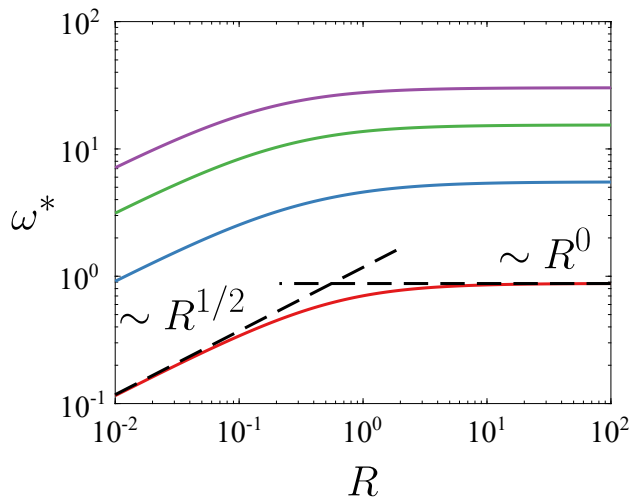

FIGURE 3. First four natural frequencies in a quiescent fluid as a function of mass ratio. Asymptotic behaviour overlaid.

Together, the results briefly shown here underline the importance of specifying the dynamical regime of the system, in particular the mass ratio $R$. All of our results will be for $R = 0.01$, and we expect our conclusions to hold for low mass ratios ($R \lesssim 0.1$).

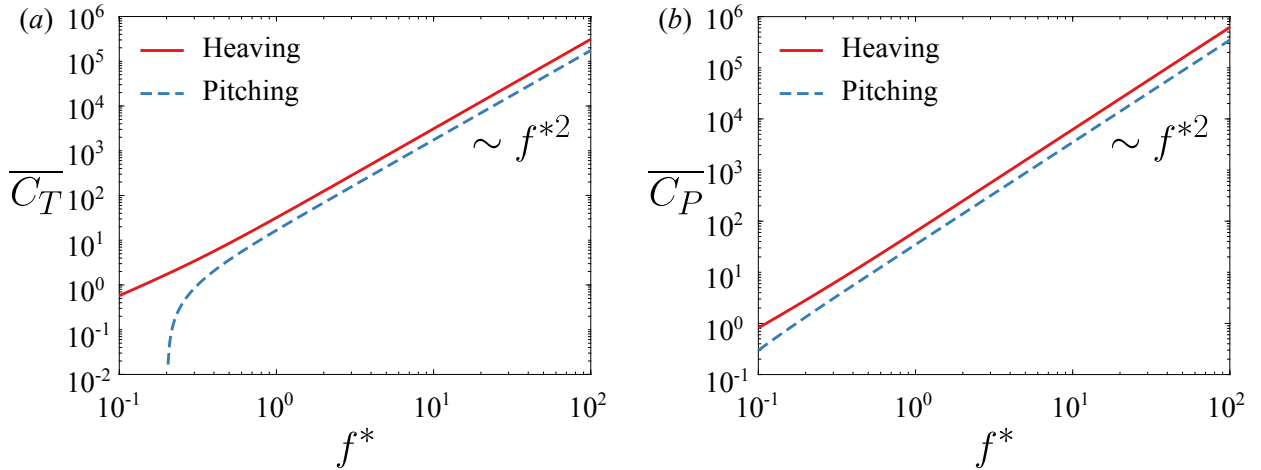

FIGURE 4. (a) Mean thrust coefficient and (b) mean power coefficient as a function of reduced frequency $f^*$ for a heaving (red) and pitching (blue) rigid plate. Asymptotic behaviour included. At low $f^*$, a pitching rigid plate produces drag in the mean.

## 4. Inviscid results

Here, we present our results on the kinematics and propulsive characteristics of uniformly flexible swimmers. Since our interests lie in clarifying the role of resonance, we limit ourselves to purely heaving and purely pitching plates; allowing simultaneous heaving and pitching would add two parameters, and would potentially dilute our results on the role of resonance. Given our interests, it also makes sense to present results for flexible plates relative to rigid plates. For example, we will present the mean thrust that a flexible plate produces relative to the mean thrust that an otherwise identical rigid plate produces. We therefore begin by briefly reviewing the results for rigid plates.

### 4.1. Propulsive characteristics of rigid swimmers

The linear inviscid theory for sinusoidally heaving and pitching rigid plates was developed in Theodorsen (1935), and extended in Garrick (1936) to provide results on the propulsive characteristics of such plates. The mean thrust produced, as well as the mean power needed to produce the mean thrust, are shown in figure 4 as a function of the reduced frequency for the amplitudes in table 2. At high reduced frequencies, the mean thrust coefficient varies as $f^{*2}$ for both heaving and pitching plates. At low reduced frequencies, the mean thrust coefficient varies sub-quadratically for a heaving plate, and super-quadratically for a pitching plate, for the reduced frequencies shown here. Note that a heaving plate always produces net thrust in the mean, whereas a pitching plate produces net drag in the mean for $f^* < 0.202$. The story is much the same for the mean power coefficient. At high reduced frequencies, the mean power coefficient varies as $f^{*2}$ for both heaving and pitching plates. At low reduced frequencies, the mean power coefficient varies sub-quadratically for a heaving plate, and super-quadratically for a pitching plate, for the reduced frequencies shown here. The power input for a heaving plate is always positive in the mean, whereas the power input for a pitching plate is negative in the mean for $f^* < 0.013$. As $f^* \to 0$, $\overline{C_T} \to 0$ and $\overline{C_P} \to 0$ for both heaving and pitching plates.

Given the mean thrust and power, we may calculate the efficiency, shown in figure 5 as a function of the reduced frequency. At high reduced frequencies, $\eta \to 0.5$ for both heaving and pitching plates. At low reduced frequencies, the efficiencies for heaving and pitching plates diverge. For a heaving plate, the efficiency increases as the reduced frequency decreases, with $\eta \to 1$ as $f^* \to 0$. For a pitching plate, the efficiency becomes negative since a pitching plate produces net drag at low reduced frequencies. Note that because of how the efficiency is defined, there is a vertical asymptote where the mean power coefficient is zero, at $f^* \approx 0.013$. To the left of this asymptote the efficiency is positive since both the mean thrust and power coefficients are negative, but we shall ignore any such cases since we are only interested in thrust-producing plates.

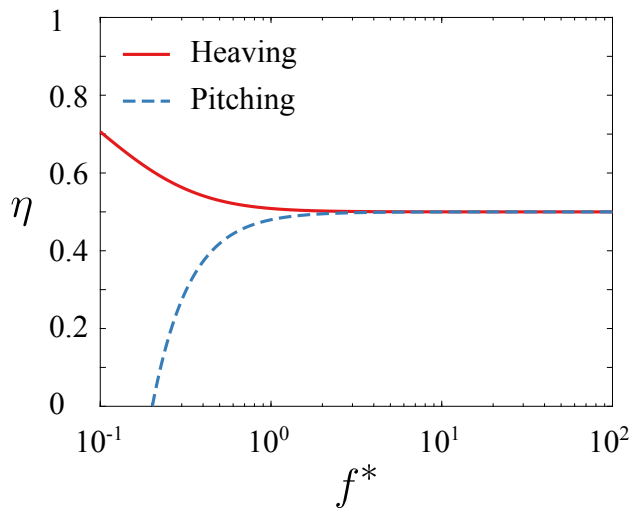

FIGURE 5. Efficiency as a function of reduced frequency $f^*$ for a heaving (red) and pitching (blue) rigid plate. At low $f^*$, a pitching rigid plate produces drag in the mean, hence its efficiency is negative.

Having briefly reviewed the propulsive characteristics of rigid plates in the linear inviscid regime, we now move on to uniformly flexible plates. It is worth bearing in mind how the thrust, power, and efficiency vary with reduced frequency for rigid heaving and pitching plates when we present the results for flexible heaving and pitching plates.

### 4.2. *Propulsive characteristics of flexible swimmers*

We begin by considering the kinematics of the flexible plate actuated sinusoidally at its leading edge. For our purposes, it is sufficient to look at the deflection at a single point along the length of the plate, which we choose to be the trailing edge. The amplitude of the trailing edge deflection is shown in figure 6. More specifically, we have plotted the logarithm of the ratio of the trailing edge amplitude of a flexible plate to the trailing edge amplitude of an otherwise identical rigid plate. The dashed white lines indicate where the flexible plate has the same trailing edge amplitude as the rigid plate. For both heaving and pitching plates, we see ridges of local maxima in trailing edge amplitude in the stiffness-frequency plane. For a given ridge, its reduced frequency increases with the nondimensional stiffness.

We suspect that the locations of the local maxima of trailing edge amplitude correspond to resonances in the system. To verify our suspicion, we formally calculate the first ten pairs of quiescent eigenvalues of the coupled fluid-structure system, that is, the eigenvalues of a clamped plate in an otherwise quiescent fluid. (Formally, by quiescent we mean in the limit where the bending velocity is large compared to the fluid velocity.) We have re-plotted the trailing edge amplitudes from figure 6 in figure 7, with the imaginary parts of the eigenvalues overlaid (and re-scaled to match the nondimensionalization employed in the plots). Indeed, the local maxima in trailing edge amplitude align with the quiescent natural frequencies of the system. The alignment is not as good when both the reduced frequency and nondimensional stiffness are low; we leave this point aside now but will revisit it later. It can be easily shown that the quiescent natural frequencies scale as $f^* \sim S^{1/2}$.

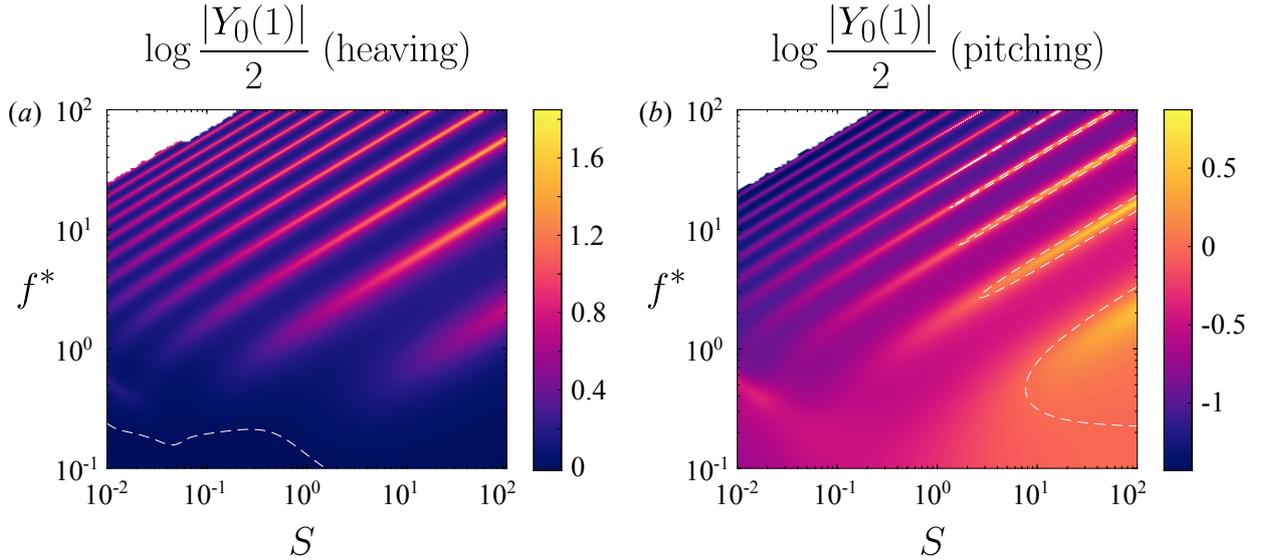

FIGURE 6. Trailing edge amplitude as a function of reduced frequency $f^*$ and stiffness ratio $S$ for a (a) heaving and (b) pitching plate with $R = 0.01$ relative to that of an equivalent rigid plate. Dashed white lines indicate where the flexible plate has the same trailing edge amplitude as the equivalent rigid plate. Under-resolved areas have been whited out.

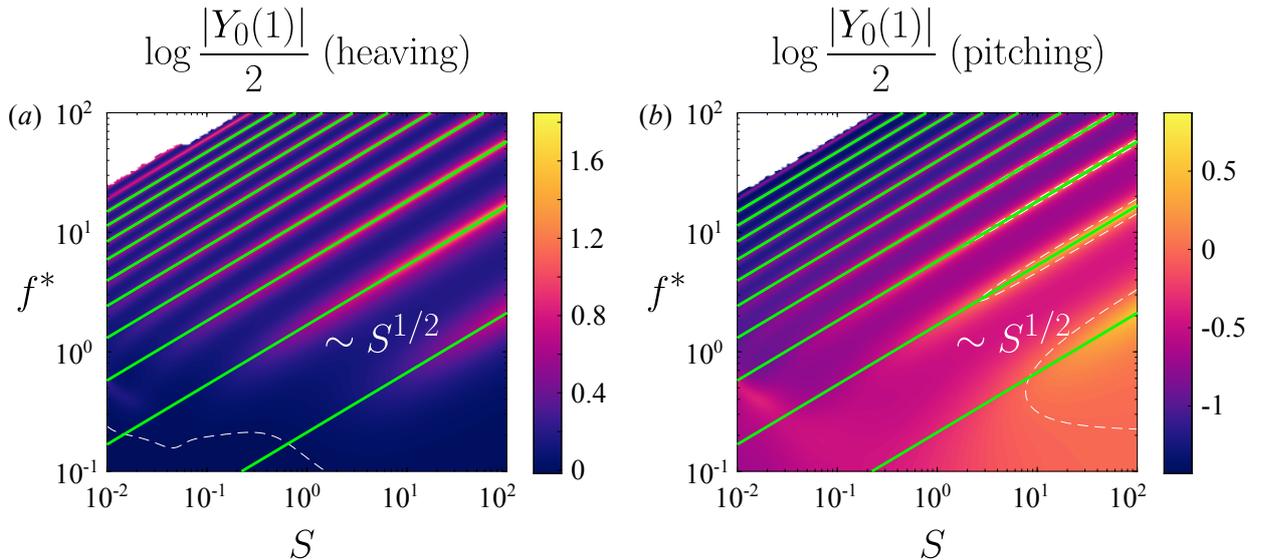

FIGURE 7. Same as in figure 6, but with quiescent natural frequencies overlaid as green lines.

To explain why the local maxima in trailing edge amplitude occur when the system is actuated at its natural frequencies, we turn to the transfer function from actuation to trailing edge deflection. Recall that the transfer function of a linear input-output system is a function $G(s)$, where $s$ is a complex number, such that the response to an input of the form $e^{st}$ is given by $G(s)e^{st}$. Since the trailing edge deflection is just a sample of the entire deflection field, the poles of the transfer function will be the eigenvalues of the system. Generally, the eigenvalues are in the left half-plane. In figure 8, we schematically illustrate the magnitude of a simple transfer function in the complex plane. In figure 8a, a single pole of the transfer function is marked as a cross, and the contour lines show level sets of the magnitude of the transfer function in the complex plane. For a single pole $\lambda$, the transfer function is $G = 1/(s - \lambda)$, where $s = \sigma + i\omega$ is the complex variable. The magnitude of the transfer function decreases with distance from the pole in the complex plane, resulting in the circular level sets centered about the pole. Since our actuation is sinusoidal (i.e., $e^{i\omega t}$), we are specifically interested in the behaviour along the imaginary axis; this is shown in figure 8b. It is clear that maxima in the magnitude of the transfer function will occur when we actuate the system at a frequency equal to the imaginary part of an eigenvalue of the system (a "natural frequency"); in other words, maxima in the magnitude of the transfer function occur when we actuate at resonance. This will generally hold true even when the system has multiple eigenvalues, as long as they are far enough from each other.

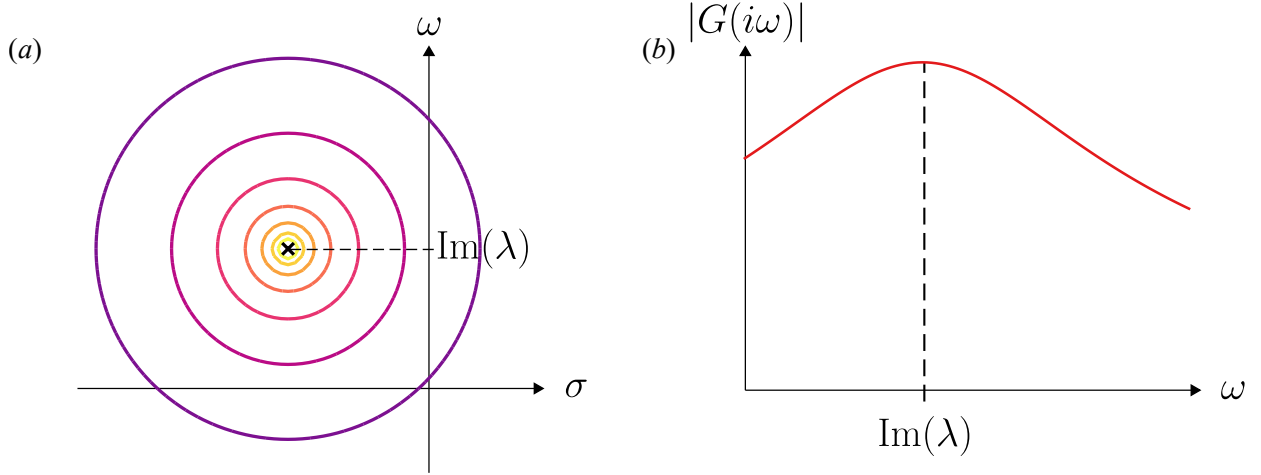

FIGURE 8. Schematic explaining resonance. (a) Level sets on the complex plane of the magnitude of a transfer function with one pole, marked with a cross. (b) Magnitude of the same transfer function evaluated on the imaginary axis.

With the swimmer's kinematics understood more or less in terms of the system's eigenvalues, we move on to its propulsive characteristics. The mean thrust and power coefficients are shown in figures 9 and 10, respectively. We have plotted the logarithm of the ratio of the mean thrust/power coefficient of a flexible plate to the mean thrust/power coefficient of an otherwise identical rigid plate to show how flexibility modifies the propulsive characteristics. The dashed white lines indicate where the flexible values match the rigid values. Regions of low reduced frequency that have negative mean thrust/power have been whited out. Just as for the trailing edge amplitude, we see ridges of local maxima in the mean thrust and power coefficients.

In figures 11 and 12, we have re-plotted the mean thrust and power coefficients, with the quiescent natural frequencies overlaid. Just as for the trailing edge amplitude, the ridges of local maxima in both mean thrust and mean power align with the quiescent natural frequencies of the system (the alignment is not as good when the reduced frequency and nondimensional stiffness are low, but we will revisit this issue later). Since the thrust and power are quadratic functions of the deflection (see (2.7) and (2.8)), we expect them to exhibit local maxima when the system is actuated at natural frequencies.

With the behaviour of the deflection, mean thrust, and mean power understood, we are left to understand the behaviour of the efficiency. The efficiency is shown in figure 13. For a heaving plate, the efficiency generally decreases with reduced frequency, just like for the rigid plate (cf. figure 5). For a pitching plate, the behaviour of the efficiency differs from the rigid case in that it increases with reduced frequency, reaches a peak, and then decreases with reduced frequency. Recall that for a rigid pitching plate, the efficiency monotonically increases with reduced frequency, not displaying any local maximum. For both heaving and pitching plates, the efficiency generally increases as the nondimensional stiffness decreases.

To isolate the effects of flexibility, we have plotted the difference in efficiency between the flexible and rigid swimmers in figure 14, with a dashed white line indicating where flexible and rigid swimmers attain the same efficiencies. We see that the flexible swimmer is broadly more efficient than the rigid swimmer (in fact, the flexible heaving plate always attains greater efficiency than the rigid heaving plate). This leads us to conclude that flexibility generally makes a swimmer more efficient, at least for low mass ratios. The mechanism for increased efficiency, however, is unclear. What about passive flexibility makes a swimmer more efficient?

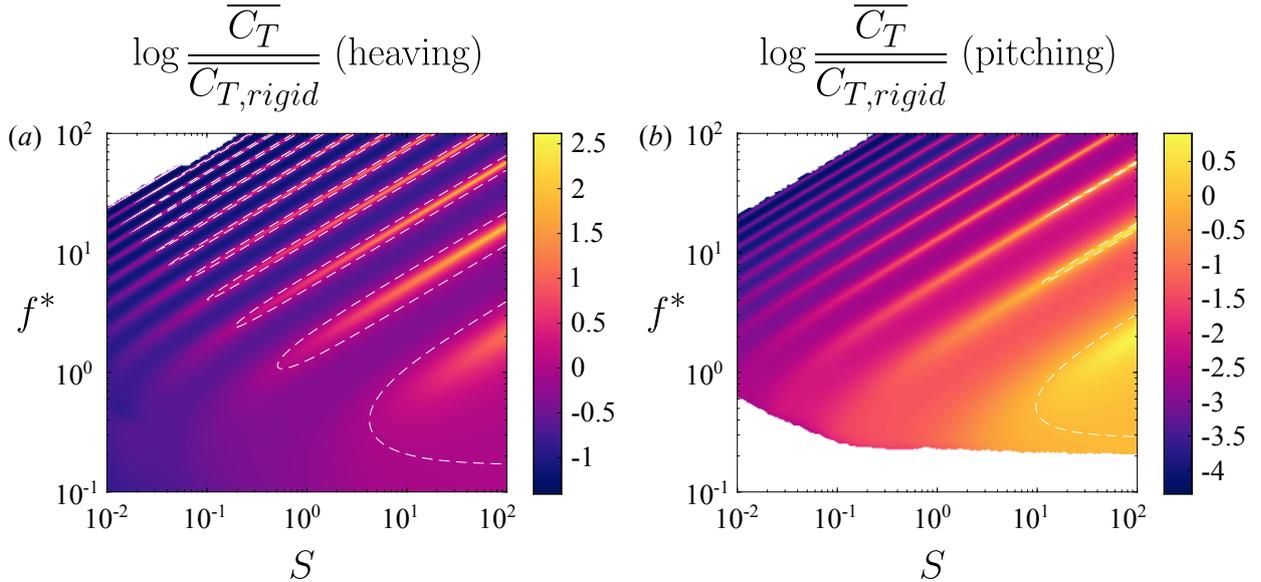

FIGURE 9. Thrust coefficient as a function of reduced frequency $f^*$ and stiffness ratio $S$ for a (a) heaving and (b) pitching plate with $R = 0.01$ relative to that of an equivalent rigid plate. Dashed white lines indicate where the flexible plate has the same thrust coefficient as the equivalent rigid plate. Under-resolved areas and areas which produce negative thrust have been whited out.

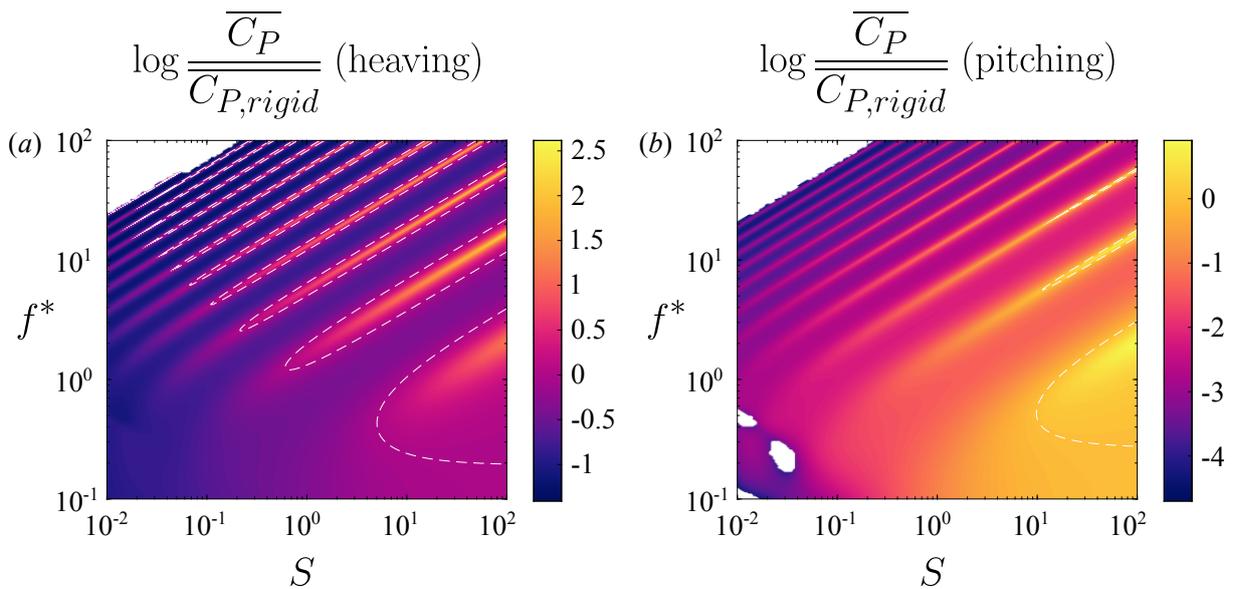

FIGURE 10. Power coefficient as a function of reduced frequency $f^*$ and stiffness ratio $S$ for a (a) heaving and (b) pitching plate with $R = 0.01$ relative to that of an equivalent rigid plate. Dashed white lines indicate where the flexible plate has the same power coefficient as the equivalent rigid plate. Under-resolved areas and areas which produce negative power input have been whited out.

It is apparent that the efficiency is not related to the quiescent natural frequencies. Whereas both mean thrust and mean power have ridges of local maxima aligned with the quiescent natural frequencies, this is not the case for the efficiency. The efficiency instead has a single broad region of high values in the stiffness-frequency plane. Elsewhere in the plane, the local maxima in thrust and power cancel each other exactly, resulting in flat efficiency; such behaviour has been previously observed in linear models of passively flexible swimmers (Alben 2008b; Moore 2014, 2017). The broad region of high efficiency is aligned with a line for which reduced frequency decreases with nondimensional stiffness, opposite the behaviour of the quiescent natural frequencies.

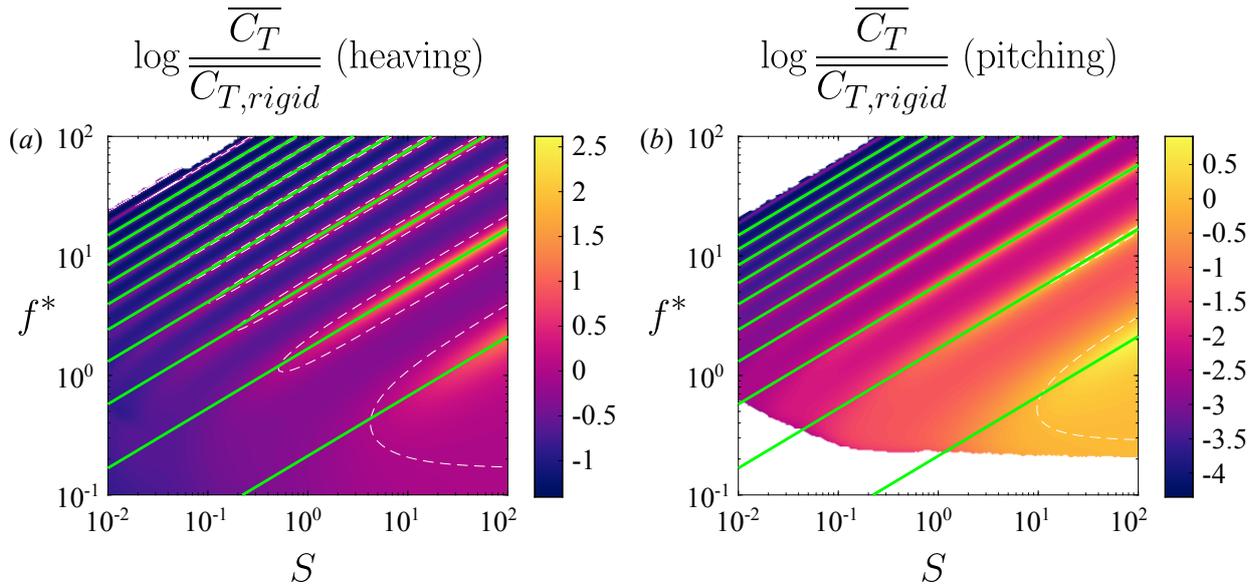

FIGURE 11. Same as in figure 9, but with quiescent natural frequencies overlaid as green lines.

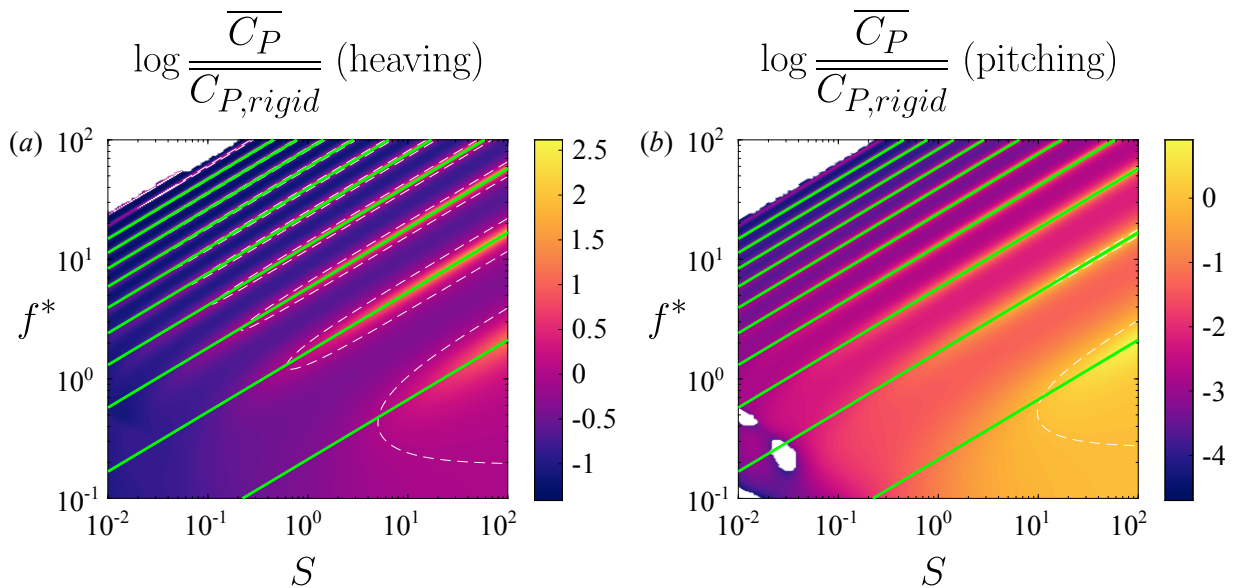

FIGURE 12. Same as in figure 10, but with quiescent natural frequencies overlaid as green lines.

### 4.3. *Fluid-structure eigenvalues and their relationship with efficiency*

While the efficiency appears to be unrelated to the quiescent natural frequencies, it may be possible that it is related to the full eigenvalues of the system. In the quiescent limit, the forces at play are the elastic forces from the plate and the added mass forces from the fluid, with lift forces being negligible. In the full problem, however, lift forces may be important. We expect the lift forces to be dominant when the reduced frequency and nondimensional stiffness are low, which is where the region of high efficiency is, and where the behaviours of the trailing edge amplitude, mean thrust, and mean power deviate from the behaviour of the quiescent eigenvalues.

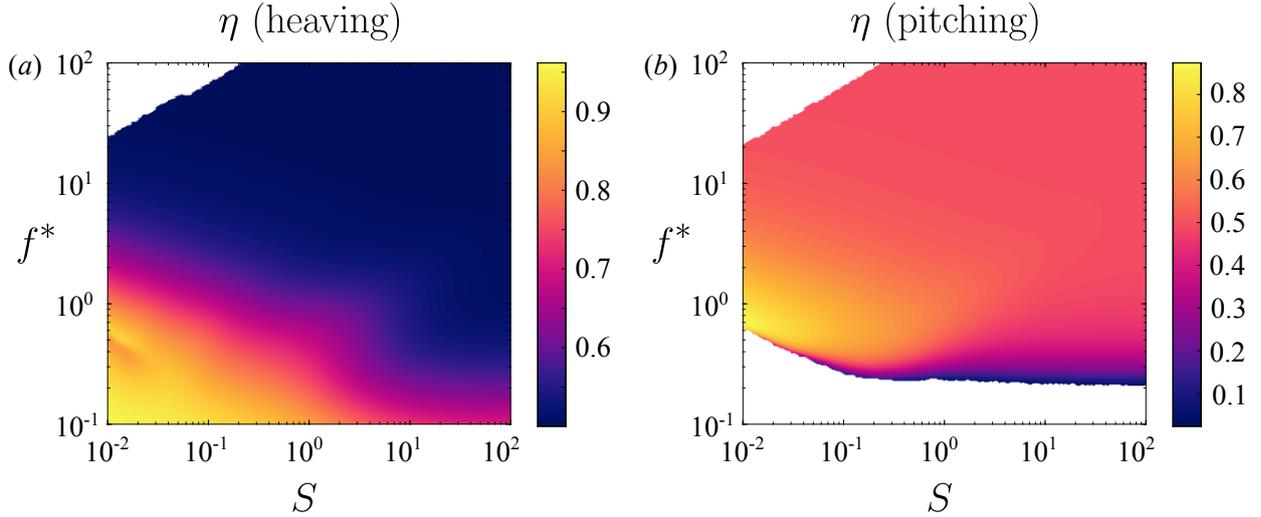

FIGURE 13. Efficiency as a function of reduced frequency $f^*$ and stiffness ratio $S$ for a (a) heaving and (b) pitching plate with $R = 0.01$. Under-resolved areas and areas with negative efficiency have been whited out.

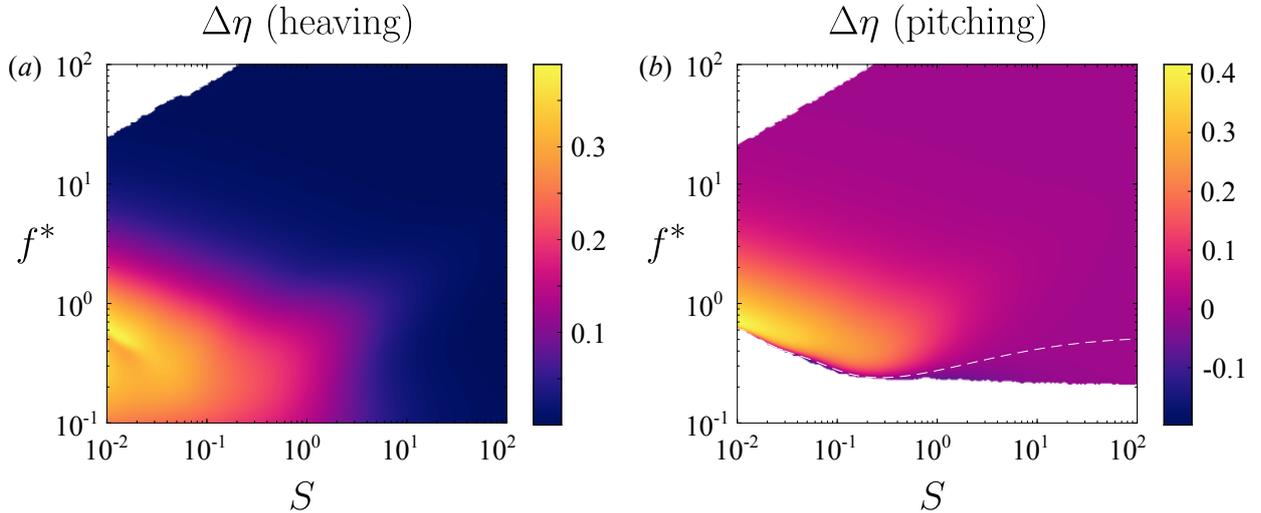

FIGURE 14. Efficiency as a function of reduced frequency $f^*$ and stiffness ratio $S$ for a (a) heaving and (b) pitching plate with $R = 0.01$ relative to that of an equivalent rigid plate. Dashed white lines indicate where the flexible plate has the same efficiency as the equivalent rigid plate. Under-resolved areas and areas with negative efficiency have been whited out.

In figure 15, we again show the difference in efficiency between flexible and rigid swimmers, but now with the full natural frequencies overlaid. When the reduced frequency and nondimensional stiffness are high, the natural frequencies match closely with the quiescent natural frequencies, as expected. For low values of the reduced frequency and nondimensional stiffness, lift forces become important and affect the natural frequencies, causing them to deviate from their quiescent behaviour. We even see the emergence of branches for which the reduced frequencies increase as the nondimensional stiffness decreases, counter to our intuition for flexible plates. The region of high efficiency is aligned with the counterintuitive branch of natural frequencies, leading us to suspect that this strange branch may be responsible for high efficiency; we therefore find it paramount to understand the behaviour of the eigenvalues of the system.

In figure 16, we trace the first three eigenvalue pairs of the full coupled fluid-structure system as the nondimensional stiffness decreases. Note that the eigenvalues are solutions of a nonlinear eigenvalue problem, so eigenvalues may appear and disappear. Also note that the imaginary parts of the eigenvalues in figure 16b are greater than those in figure 15 by a factor of $\pi$ because of how we have chosen to define $f^*$. In the following description of the eigenvalues, we begin at large stiffness ratio $S$ and describe how the eigenvalues change as we decrease $S$, since the eigenvalues essentially behave as those for an Euler-Bernoulli beam in vacuo for large $S$.

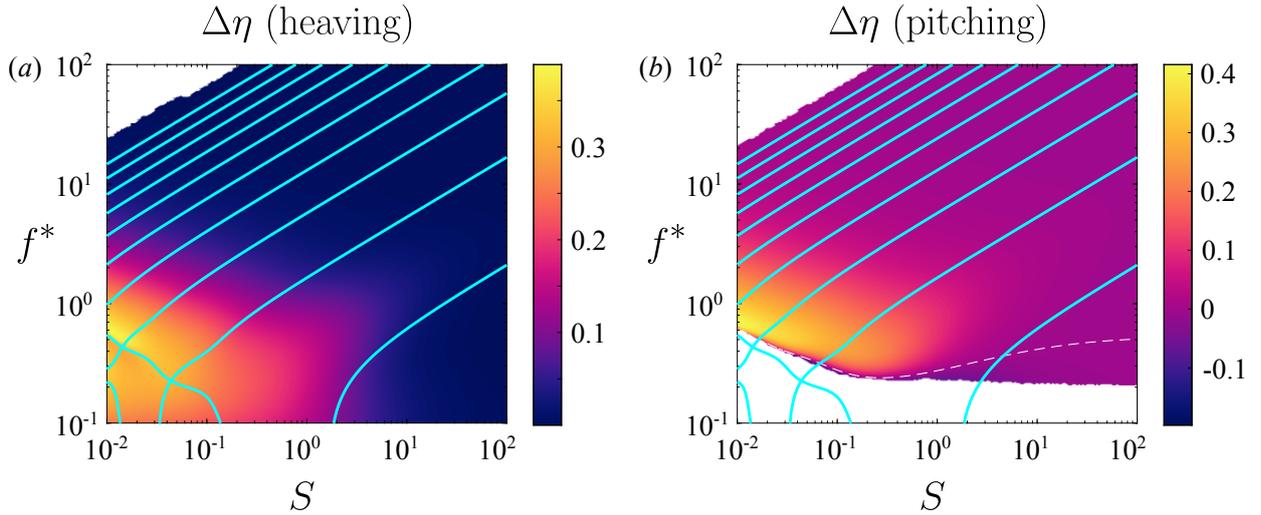

FIGURE 15. Same as in figure 14, but with natural frequencies overlaid as cyan lines.

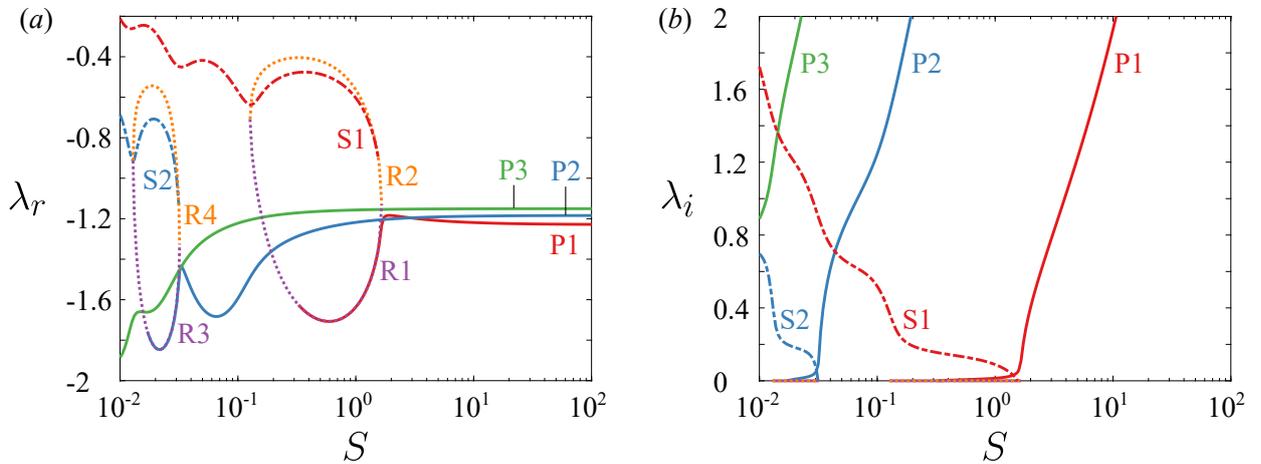

FIGURE 16. First few eigenvalues of the system as a function of stiffness ratio $S$: (a) real parts; and (b) imaginary parts.

As the stiffness ratio decreases, the first three eigenvalues behave as expected: the imaginary parts decrease, and the real parts do not change. We shall refer to these as primary eigenvalues, and label them P1, P2, and P3. As $S$ further decreases, the behaviour of P1 changes: its real part first increases a bit, then decreases dramatically, and then begins to loop up; its imaginary part first decreases more quickly, then decreases substantially more slowly, until it finally decays to zero. At this point ($S = 0.327$), P1 merges with one of the two real eigenvalues that have appeared, labelled R1 and R2. The two real eigenvalues appear when $S = 1.658$, shortly after the behaviour of P1 changes, and they merge and disappear when $S = 0.127$. Just after the two real eigenvalues appear, a new conjugate pair, labelled S1, appears when $S = 1.549$. We refer to this eigenvalue as a secondary eigenvalue because it essentially replaces the primary eigenvalue P1. We summarize the behaviour as follows: the original primary eigenvalue, P1, has decreasing imaginary part until it becomes purely real. In this time, a pair of real eigenvalues, R1 and R2, appear. P1 and R1 merge when P1 becomes real, and they eventually disappear, along with R2, when they all collide. A new conjugate pair of secondary eigenvalues, S1, appears as well, and both its real and imaginary parts increase as $S$ decreases. The second primary eigenvalue P2 essentially demonstrates the same behaviour, and we hazard a guess that P3 shows the beginnings of the same behaviour.

What physical mechanism is at the root of the observed behaviour in the eigenvalues? It should be clear that P1 is an Euler-Bernoulli mode, since it essentially displays the behaviour of an eigenvalue of an Euler-Bernoulli beam in vacuo. To be more precise, the behaviour of P1 is dominated by elastic and added mass forces, leading to Euler-Bernoulli type behaviour. S1, on the other hand, is a flutter mode. S1 emerges when the stiffness ratio is low, and so its behaviour is dominated by lift and added mass forces. Both the real and imaginary parts of S1 increase as the stiffness ratio decreases, characteristic of a flutter mode. The stiffness ratio can also be thought of as the inverse of a reduced flow velocity, as in Eloy *et al.* (2007), whereby increasing the reduced flow velocity leads to a flutter instability. If we decreased $S$ even further, S1 would eventually become unstable. When the nondimensional stiffness is $O(1)$, P1, S1, R1, and R2 simultaneously exist. For such values of the nondimensional stiffness, all three types of forces— elastic, lift, and added mass— are non-negligible. The importance of all three types of forces explains why all three modes— Euler-Bernoulli, flutter, and divergence (R1 and R2)— simultaneously exist.

The same emergence and disappearance of modes occurs for the higher-order modes as well, but at lower values of the stiffness ratio. The change in behaviour occurs at lower values of $S$ because the higher-order modes have shorter wavelengths (see Alben 2008*a*, for example), significantly increasing the $Y_{xxxx}$ term in (2.3), thereby significantly increasing the magnitude of the elastic forces. We therefore expect the elastic forces to become dominated by lift forces at much lower values of $S$ for the higher-order modes.

As the nondimensional stiffness decreases, the eigenvalues of the system come closer together in the complex plane. When multiple eigenvalues are relatively close to each other, our notion of resonance, schematically illustrated in figure 8, becomes muddied. In figure 17, we schematically illustrate the magnitude of a transfer function with *multiple* poles that are relatively close to each other. In figure 17a, the poles of the transfer function are marked as crosses, and the contour lines show level sets of the magnitude of the transfer function in the complex plane. Because the poles are close to each other, the level sets are no longer simple circles. Since our actuation is sinusoidal, we are specifically interested in the behaviour along the imaginary axis; this is shown in figure 17b. Because the poles are close to each other, there are no longer local maxima when the system is actuated at one of its natural frequencies; instead, there is a broad response across the range of natural frequencies. In our example, there is a single local maximum despite there being four poles. Moreover, the local maximum does not occur at any of the natural frequencies of the system, it occurs between the imaginary parts of $\lambda_3$ and $\lambda_4$. This schematic explains why the ridges of local maxima in trailing edge amplitude, mean thrust, and mean power broaden and smear together as the reduced frequency and nondimensional stiffness become small (see figures 6, 9, and 10).

With a good understanding of the eigenvalues of the system, we may now interpret the behaviour of the efficiency in light of the behaviour of the eigenvalues. Specifically, we want to understand the difference in efficiency between flexible and rigid swimmers (i.e. figures 14 and 15). Broadly speaking, we expect a flexible swimmer to be more efficient than a rigid one for a simple reason: as a flexible swimmer moves through the fluid, its body deforms in response to the forcing from the fluid, so it does not need to fight against the fluid as much as a non-deforming rigid swimmer does. A flexible swimmer therefore expends less energy in driving its motion than a rigid swimmer does. This effect becomes more pronounced as the elastic forces weaken relative to the lift forces— as the swimmer becomes flimsier. As previously discussed, the elastic forces weaken relative to the lift forces as the nondimensional stiffness $S$ decreases. The elastic forces also become relatively weaker when the frequency of actuation is decreased. As we can see from the eigenvalues in figure 15, a lower reduced frequency will excite lower-order modes. The lower-order modes have longer wavelengths, and therefore relatively weaker elastic forces. To summarize, decreasing $S$ and $f^*$ weakens the elastic forces in the swimmer, thereby weakening its ability to resist the fluid, lowering the power needed to drive its motion, and making it more efficient.

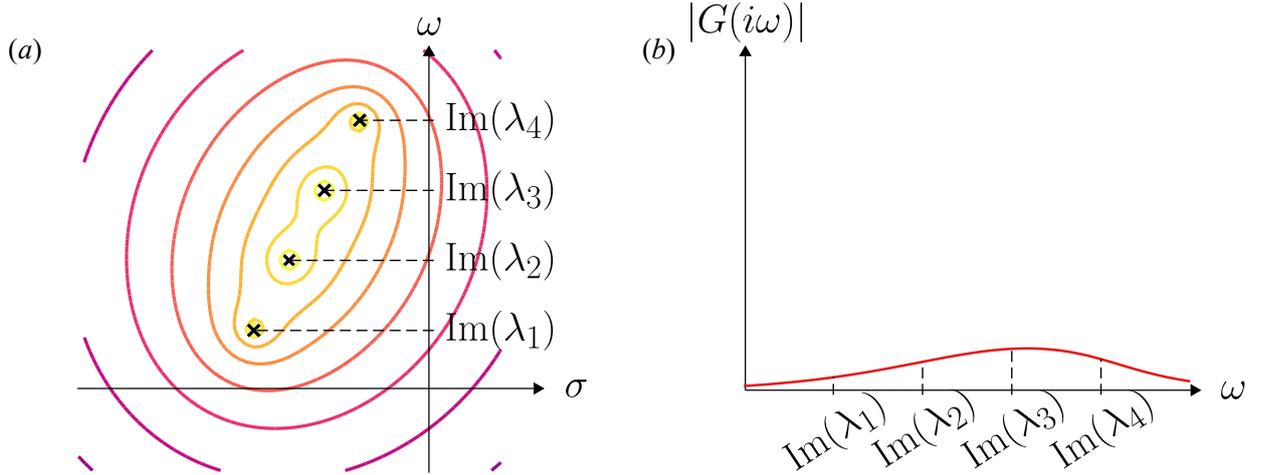

FIGURE 17. Our simple notion of resonance becomes unclear when multiple poles are relatively close. (a) Level sets on the complex plane of the magnitude of a transfer function with four poles, marked with crosses. (b) Magnitude of the same transfer function evaluated on the imaginary axis.

This is not the complete picture, however. As we can see in figure 15, in the lower left region there are areas where decreasing $S$ and *increasing* $f^*$ improves the efficiency, counter to our previous argument. This behaviour can be understood in terms of the changing behaviour of the eigenvalues of the system in that region. When $S$ becomes small enough, the primary eigenvalue is essentially replaced by the secondary eigenvalue. Recall that the primary eigenvalue corresponds to an Euler-Bernoulli mode, and the secondary eigenvalue corresponds to a flutter mode. Euler-Bernoulli modes are dominated by elastic forces, while flutter modes are dominated by lift forces. Based on the previous discussion, swimmers whose composition includes flutter modes should be more efficient.

A comparison between Euler-Bernoulli mode P2 and flutter mode S1 for $S = 0.1$ is shown in figure 18, where the modes have been normalized so that their second derivatives at the leading edge are real and equal to 1 (a supplementary video is included). Qualitatively, the flutter mode looks more efficient than the Euler-Bernoulli mode, with the Euler-Bernoulli mode having a rigid fore and aft (this may be easier to see in the supplementary video). To quantify this observation, in figure 19 we have plotted the magnitudes of the modes as well as the phase between the leading edge and the deflection along the chord for the two modes, normalized as before. The deflection of flutter mode S1 is greater than that of Euler-Bernoulli mode P2 along the entire chord. The phase is flat for a large portion of the fore of Euler-Bernoulli mode P2, indicating that it moves rigidly. The phase also flattens out towards the aft, indicating that it too is nearly rigid. In contrast, flutter mode S2 has a nearly linearly decreasing phase. A front-to-back travelling wave would have a linearly decreasing phase, so flutter mode S1 essentially behaves as a travelling wave (with spatially-varying amplitude). As shown in Wu (1961), travelling wave kinematics can be quite efficient. The emergence of the flutter modes as $S$ decreases leads to travelling wave kinematics in the actuated system. For a value of $S$ for which a flutter mode exists, the phase of the deflection decreases nearly linearly when the system is actuated at a low frequency, indicating that the kinematics are nearly a travelling wave. As the frequency of actuation is increased, the phase behaves less linearly, instead alternating between relatively flat and steep behaviours; the degradation of the travelling wave kinematics is more severe as the frequency of actuation is increased. The behaviour at low frequencies is therefore dominated by the flutter modes, while the Euler-Bernoulli modes become dominant at higher frequencies. The frequency at which the travelling wave kinematics degrade increases as $S$ decreases, coinciding with the frequency at which the efficiency degrades, and with the behaviour of the imaginary parts of the flutter eigenvalues. We may therefore reasonably conclude that the emergence of the flutter modes as $S$ decreases makes the swimmer more efficient.

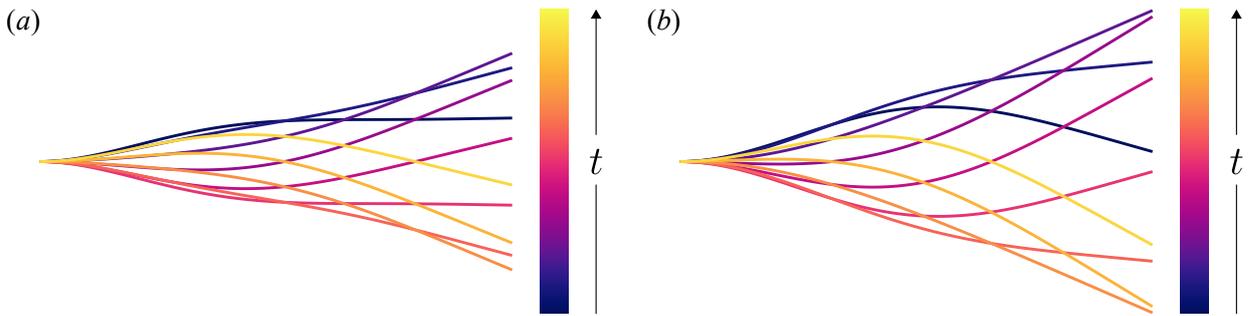

FIGURE 18. Ten snapshots, evenly spaced in time, of (a) Euler-Bernoulli mode P2 and (b) flutter mode S1 for $S = 0.1$ comprising one period of motion. The modes have been normalized so that their second derivatives at the leading edge are real and equal to 1. Supplementary videos are included.

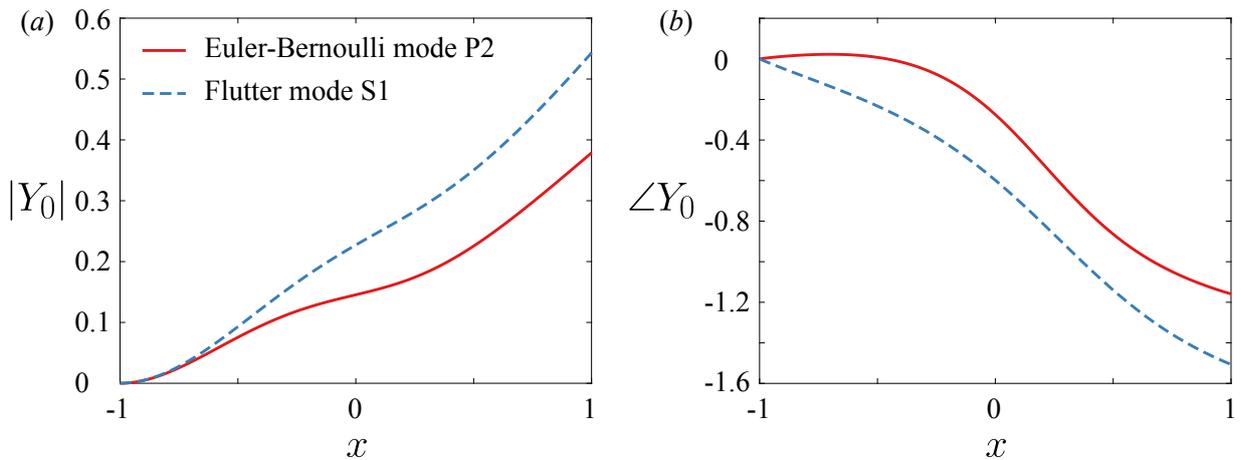

FIGURE 19. (a) Magnitude and (b) phase in radians of the deflection along the chord for Euler-Bernoulli mode P2 (red) and flutter mode S1 (blue), for $S = 0.1$. The modes have been normalized so that their second derivatives at the leading edge are real and equal to 1.

As a final note, we point out that increases in efficiency are often intertwined with decreases in thrust. This is apparent when comparing the plots of mean thrust with the plots of efficiency (figures 9 and 14, respectively). To generate large thrust, the swimmer needs to be able to push against the fluid. To be efficient, however, the swimmer needs to be compliant to the fluid. A limiting case of this is when the body of the swimmer takes the form of a front-to-back travelling wave. As the wave velocity approaches the freestream velocity, the thrust vanishes, the efficiency approaches unity, and the swimmer merely travels along a sinusoidal path fixed in space (Wu 1961). We must be mindful of regions of low thrust, especially in the presence of drag, as we shall explore in the next section.

## 5. Finite Reynolds number effects

Recently, the effects of streamwise drag on efficiency have come to be appreciated, at least for rigid swimmers (Floryan *et al.* 2017). Drag can create peaks in efficiency and can make the efficiency quite sensitive to changes in the system, as also suggested in Moore (2014). Here, we consider how streamwise drag due to a finite Reynolds number affects the system.

The presence of drag in our system does not change it much. The kinematics will not change, so the trailing edge amplitude remains unchanged. The net thrust produced decreases uniformly across the stiffness-frequency plane, leaving the picture qualitatively the same. The power consumption also does not change. The efficiency, however, will change. Whereas before there were no local maxima in efficiency, the addition of an offset drag to the system will spur the emergence of ridges of local maxima in efficiency, just like the ones previously described for the trailing edge amplitude, mean thrust, and mean power.

The ridges of local maxima in efficiency caused by the addition of an offset drag can be understood in a simple way. We will consider a simplified picture of resonance in our system. Suppose we actuate the inviscid system at a non-resonant frequency, resulting in mean thrust coefficient $\overline{C_{T0}}$, mean power coefficient $\overline{C_{P0}}$, and efficiency $\eta_0 = \overline{C_{T0}}/\overline{C_{P0}}$. If we change the frequency of actuation to a resonant one, our previous results show that the mean thrust coefficient, power coefficient, and efficiency will become

$$\overline{C_{T1}} = a\overline{C_{T0}}, \quad \overline{C_{P1}} = a\overline{C_{P0}}, \quad \eta_1 = \frac{\overline{C_{T1}}}{\overline{C_{P1}}} = \frac{a\overline{C_{T0}}}{a\overline{C_{P0}}} = \frac{\overline{C_{T0}}}{\overline{C_{P0}}} = \eta_0, \qquad (5.1)$$

where $a > 1$. We see that resonance does not alter the efficiency when there is no drag.

Now consider adding streamwise drag to the system. The baseline mean thrust changes by an offset, and the mean power does not change. The baseline efficiency is then

$$\eta_0 = \frac{\overline{C_{T0}} - C_D}{\overline{C_{P0}}} = \frac{\overline{C_{T0}}}{\overline{C_{P0}}} - \frac{C_D}{\overline{C_{P0}}}, \qquad (5.2)$$

where $C_D$ is the drag coefficient. When we actuate at resonance, the mean thrust and mean power increase as before, and the drag does not change. The efficiency becomes

$$\eta_1 = \frac{a\overline{C_{T0}} - C_D}{a\overline{C_{P0}}} = \eta_0 + \frac{a-1}{a}\frac{C_D}{\overline{C_{P0}}} > \eta_0. \qquad (5.3)$$

We see that the addition of streamwise drag to the system causes local maxima in efficiency when the system is actuated at a natural frequency. This effect should be robust to the source of drag. Note that this effect depends on how strongly resonance affects the system (the value of $a$), and on how strong the drag is ($C_D/\overline{C_{P0}}$). We demonstrate this effect in figure 20, where we show the efficiency for $C_D = 0.1$ (Floryan *et al.* 2017). Since the effect depends on the ratio $C_D/\overline{C_{P0}}$, just for this plot we have changed the amplitudes to $h_0 = 0.2$ and $\theta_0 = 0.1$. Indeed, we see ridges of local maxima in efficiency which align with the natural frequencies. We also note that the addition of streamwise drag has pushed the thrust-drag transition to significantly higher values of the reduced frequency; this underscores the importance of streamwise drag for swimmers.

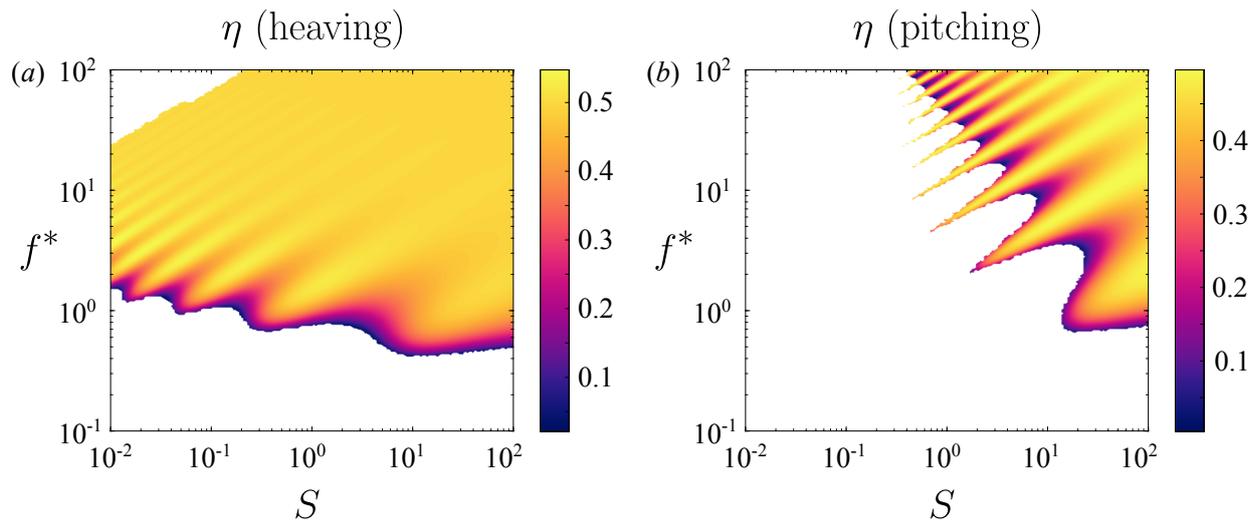

FIGURE 20. Efficiency as a function of reduced frequency $f^*$ and stiffness ratio $S$ for a (a) heaving and (b) pitching plate with $R = 0.01$ with additional drag. Under-resolved areas and areas with negative efficiency have been whited out.

Since any real system will have some drag, resonant peaks in efficiency should be present. We offer our simple explanation as a reason for the existence of resonant peaks in efficiency observed in the literature, modulo nonlinear effects. Since our analysis is linear, the aforementioned effect of streamwise drag on the efficiency of the system is present at first order, and we therefore expect it to be robust to nonlinear effects.

## 6. Conclusions

In this work, we studied a linear inviscid model of a passively flexible swimmer, valid for small-amplitude, low-frequency motions where there is no separation. The frequencies of actuation and stiffness ratios we considered spanned a large range, while the mass ratio was mostly fixed to a low value representative of swimmers. A short set of results for which we varied the mass ratio indicates that there exist qualitative differences between flappers with low mass ratios (swimmers) and those with mass ratios of order unity and higher (fliers). The results presented in this work are therefore applicable to swimmers, and care should be taken in extending the results to fliers.

We presented results showing how the trailing edge deflection, thrust coefficient, power coefficient, and efficiency vary in the stiffness-frequency plane. The trailing edge deflection, thrust coefficient, and power coefficient showed sharp ridges of resonant behaviour for reduced frequencies $f^* > 1$ and stiffness ratios $S > 1$. In this region, the locations of the resonant peaks were well-predicted by the imaginary parts of the quiescent eigenvalues of the system. For $f^* < 1$ and $S < 1$, however, the resonant peaks smeared together. The efficiency, on the other hand, did not show resonant peaks anywhere in the stiffness-frequency plane, instead showing a broad region of high values for $f^* < 1$ and $S < 1$.

Calculating the full eigenvalues and eigenfunctions of the system, we saw that the region of high efficiency coincided with the emergence of flutter modes and disappearance of Euler-Bernoulli modes. The imaginary parts of the eigenvalues of the flutter modes increase with decreasing stiffness ratio, opposite to the behaviour of the Euler-Bernoulli modes. The eigenfunctions revealed that flutter modes take on a form close to a travelling wave, whereas the Euler-Bernoulli modes have nearly rigid regions. In the actuated system, cases with high efficiency took on near-travelling wave forms, and the degradation of efficiency coincided with a degradation of the travelling wave. We may therefore reasonably conclude that the emergence of the flutter modes as $S$ decreases makes the swimmer more efficient.

Lastly, we considered the effects of a finite Reynolds number in the form of streamwise drag. Streamwise drag added an offset drag to the system, which created resonant peaks in the efficiency that are not present in the inviscid system. Since any real system will have some streamwise drag, resonant peaks should be present. We offer our simple explanation as a reason for the existence of resonant peaks in efficiency observed in the literature.

This work was supported by ONR Grant N00014-14-1-0533 (Program Manager R. Brizzolara). We would also like to thank Dr A. Goza for many useful discussions.

## Appendix A. Method of solution

Consider the case where the imposed leading edge motion is sinusoidal in time with dimensionless angular frequency $\sigma = \pi L f/U$, where $f$ is the dimensional frequency in Hz. We may then decompose the deflection into a product of temporal and spatial terms, with the temporal component being sinusoidal and the spatial component represented by a Chebyshev series:

$$\left.\begin{aligned} Y(x,t) &= e^{j\sigma t} Y_0(x), \\ Y_0(x) &= \frac{1}{2}\beta_0 + \sum_{k=1}^{\infty} \beta_k T_k(x), \end{aligned}\right\} \quad (A\,1)$$

where $j = \sqrt{-1}$, the real part in j should be taken when evaluating the deflection, and $T_k(x) = \cos(k \arccos x)$ is the Chebyshev polynomial of degree $k$. For a given deflection $Y$ of this form, the solution to the flow is given in Wu (1961); we repeat the basics of that analysis in the proceeding text.

Represent 2D physical space $(x,y)$ by the complex plane $z = x + iy$, where $i = \sqrt{-1}$ but $ij \neq -1$. There exists a complex potential $F(z,t) = \phi(z,t) + i\psi(z,t)$, with $\phi$ and $\psi$ harmonic conjugates, that is analytic in $z$ and related to the complex velocity $w = u - iv$ through the momentum equation by

$$\frac{\partial F}{\partial z} = \frac{\partial w}{\partial t} + \frac{\partial w}{\partial z}. \quad (A\,2)$$

We use the conformal transformation

$$z = \frac{1}{2}\left(\zeta + \frac{1}{\zeta}\right) \quad (A\,3)$$

to map physical space in the $z$-plane to the exterior of the unit circle in the $\zeta$-plane. This transformation maps the plate onto the unit circle. The complex potential can be represented by a multipole expansion

$$F(\zeta,t) = \phi(\zeta,t) + i\psi(\zeta,t) = ie^{j\sigma t}\left(\frac{a_0}{\zeta+1} + \sum_{k=1}^{\infty}\frac{a_k}{\zeta^k}\right). \quad (A\,4)$$

Evaluating on the unit circle $\zeta = e^{i\theta}$ gives

$$\left. \begin{aligned} \phi(\zeta = e^{i\theta}, t) &= e^{j\sigma t}\left(\frac{1}{2}a_0 \tan\frac{\theta}{2} + \sum_{k=1}^{\infty} a_k \sin k\theta\right), \\ \psi(\zeta = e^{i\theta}, t) &= e^{j\sigma t}\left(\frac{1}{2}a_0 + \sum_{k=1}^{\infty} a_k \cos k\theta\right). \end{aligned} \right\} \quad (A\,5)$$

In physical space, on the surface of the plate we have

$$\left. \begin{aligned} \phi(z = x \pm 0i, t) &= e^{j\sigma t}\Phi^{\pm}(x) = e^{j\sigma t}\left(\pm\frac{1}{2}a_0\sqrt{\frac{1-x}{1+x}} \pm \sum_{k=1}^{\infty} a_k \sin k\theta\right), \\ \psi(z = x \pm 0i, t) &= e^{j\sigma t}\Psi(x) = e^{j\sigma t}\left(\frac{1}{2}a_0 + \sum_{k=1}^{\infty} a_k T_k(x)\right), \end{aligned} \right\} \quad (A\,6)$$

where we have used $x = \cos\theta$. $\psi$ has equal values on the top and bottom since it is even in $\theta$, whereas $\phi$ is odd in $\theta$ and thus has a discontinuity in physical space.

The no-penetration condition can be written as

$$\frac{\partial \psi}{\partial x}\Big|_{y=0} = -\left(\frac{\partial}{\partial t} + \frac{\partial}{\partial x}\right)^2 Y, \quad (A\,7)$$

which simplifies to

$$D\Psi = -(j\sigma + D)^2 Y_0, \quad (A\,8)$$

where $D = d/dx$. Given $Y_0$, this equation allows us to solve for all $a_k$ except $a_0$. To solve for $a_0$, we begin by writing the vertical velocity on the surface of the plate as

$$v(z = x + 0i, t) = e^{j\sigma t}V(x) = e^{j\sigma t}\left(\frac{1}{2}V_0 + \sum_{k=1}^{\infty} V_k T_k(x)\right). \quad (A\,9)$$

The no-penetration condition can then be written as

$$V = (j\sigma + D)Y_0. \quad (A\,10)$$

The coefficient $a_0$ is given by

$$a_0 = -C(j\sigma)(V_0 + V_1) + V_1, \quad (A\,11)$$

where

$$C(j\sigma) = \frac{K_1(j\sigma)}{K_0(j\sigma) + K_1(j\sigma)} \quad (A\,12)$$

is the Theodorsen function, and $K_\nu$ is the modified Bessel function of the second kind of order $\nu$. The expression for $a_0$ is derived in Wu (1961).

With all of the $a_k$ known, the pressure difference across the plate can be written as

$$\Delta p(x, t) = e^{j\sigma t}P_0(x) = e^{j\sigma t}\left(a_0\sqrt{\frac{1-x}{1+x}} + 2\sum_{k=1}^{\infty} a_k \sin k\theta\right). \quad (A\,13)$$

We note that the pressure difference depends linearly on the deflection $Y_0$.

Altogether, given the deflection $Y_0$, we may calculate the coefficients $a_k$. The coefficients $a_k$ are used to calculate the pressure difference across the plate, which alters the deflection of the plate via (2.3). The coupled fluid-structure problem must be solved numerically.

A.1. *Numerical method*

Substituting the Chebyshev series (A 1) into the Euler-Bernoulli equation (2.3) gives a fourth-order differential equation for $Y_0$:

$$-2\sigma^2 R Y_0 + \frac{2}{3} S D^4 Y_0 = P_0. \quad (A\,14)$$

The corresponding boundary conditions (2.6) are re-written as

$$Y_0(-1) = h_0, \quad Y_{0,x}(-1) = \theta_0, \quad Y_{0,xx}(1) = 0, \quad Y_{0,xxx}(1) = 0, \quad (A\,15)$$

where $h_0$ and $\theta_0$ are the heaving and pitching amplitudes at the leading edge, respectively. We re-iterate that the pressure difference across the plate $P_0$ is a linear function of the deflection $Y_0$, and so (A 14)–(A 15) give a linear, homogemeous boundary value problem for $Y_0$. When solving for the deflection $Y_0$, all infinite series are truncated to the upper limit $N$.

The numerical method to solve the boundary value problem is given in Moore (2017). The method is a pseudo-spectral Chebyshev scheme that uses Gauss-Chebyshev points. The method is fast ($O(N \log N)$) and accurate, avoiding errors typically encountered when using Chebyshev methods to solve high-order differential equations by pre-conditioning the system with continuous operators. Quadrature formulas for the thrust and power coefficients in (2.7) and (2.8) are also given in Moore (2017).

## Appendix B. Eigenvalues of the system

Here, we seek to determine the natural response of a flexible plate whose leading edge is held clamped in an oncoming flow (Alben 2008*a*; Michelin & Llewellyn Smith 2009; Eloy *et al.* 2007). This amounts to finding the eigenvalues and eigenvectors of the system (2.3) with homogeneous boundary conditions ($h(t) \equiv 0$ and $\theta(t) \equiv 0$). To do so, quantities that were previously written as Fourier-Chebyshev expansions (the deflection, complex potential, and velocity) are now written as Chebyshev series with time-varying coefficients. Following the preceeding analysis, we arrive at the following equations:

$$2R Y_{tt} + \frac{2}{3} S Y_{xxxx} = \Delta p, \quad (B\,1)$$

$$Y(x,t) = \frac{1}{2}\beta_0(t) + \sum_{k=1}^{\infty} \beta_k(t) T_k(x), \quad (B\,2)$$

$$\Delta p(x,t) = a_0(t)\sqrt{\frac{1-x}{1+x}} + 2\sum_{k=1}^{\infty} a_k(t)\sin k\theta, \quad (B\,3)$$

$$\sum_{k=1}^{\infty} a_k T'_k = -\frac{1}{2}\ddot{\beta}_0 - \sum_{k=1}^{\infty}\left[\ddot{\beta}_k T_k + 2\dot{\beta}_k T'_k + \beta_k T''_k\right], \quad (B\,4)$$

where a dot denotes differentiation with respect to $t$ and a prime denotes differentiation with respect to $x$.

As before, we need an additional equation to determine $a_0$. For now, we use (A 11) but treat the Theodorsen function as a constant $C$. The coefficient $a_0$ is then

$$a_0 = -C(V_0 + V_1) + V_1, \quad (B\,5)$$

where $V_k$ is the $k^{\text{th}}$ Chebyshev coefficient of the vertical velocity on the surface of the plate. The $V_k$ are obtained by evaluating the no-penetration condition (2.5):

$$\frac{1}{2}V_0 + \sum_{k=1}^{\infty} V_k T_k = \frac{1}{2}\dot{\beta}_0 + \sum_{k=1}^{\infty}\left[\dot{\beta}_k T_k + \beta_k T'_k\right]. \tag{B 6}$$

Treating $a_0$ in this manner will yield a linear eigenvalue problem. After obtaining the eigenvalues and eigenfunctions of the linear eigenvalue problem, we will use those as initial guesses for the nonlinear eigenvalue problem, which will use the full Theodorsen function. But first, we proceed with the description of the linear eigenvalue problem.

We can write the equations more compactly as follows:

$$2R\ddot{\boldsymbol{\beta}} + \frac{2}{3}S\boldsymbol{D}^4\boldsymbol{\beta} = \boldsymbol{P}, \tag{B 7}$$

$$\boldsymbol{P} = \boldsymbol{A}\boldsymbol{a}, \tag{B 8}$$

$$\boldsymbol{D}\boldsymbol{a} = -\ddot{\boldsymbol{\beta}} - 2\boldsymbol{D}\dot{\boldsymbol{\beta}} - \boldsymbol{D}^2\boldsymbol{\beta}, \tag{B 9}$$

$$\boldsymbol{V} = \dot{\boldsymbol{\beta}} + \boldsymbol{D}\boldsymbol{\beta}, \tag{B 10}$$

with (B 5) for $a_0$. In the above, $\boldsymbol{\beta}$ is a vector of the Chebyshev coefficients of the deflection $Y$, and similarly for $\boldsymbol{P}$ (pressure), $\boldsymbol{a}$ (potential), and $\boldsymbol{V}$ (vertical velocity). $\boldsymbol{P} = \boldsymbol{A}\boldsymbol{a}$ simply states that the Chebyshev coefficients of the pressure are linear combinations of the coefficients $a_k$, and $\boldsymbol{D}$ is the spectral representation of the differentiation operator.

Putting everything together, we get the following ODE:

$$2R\ddot{\boldsymbol{\beta}} + \frac{2}{3}S\boldsymbol{D}^4\boldsymbol{\beta} = \boldsymbol{A}[-\boldsymbol{D}^-\ddot{\boldsymbol{\beta}} - 2\boldsymbol{D}^-\boldsymbol{D}\dot{\boldsymbol{\beta}} + \boldsymbol{e_1}(\boldsymbol{e_2} - C\boldsymbol{e_1} - C\boldsymbol{e_2})^T\dot{\boldsymbol{\beta}}$$
$$-\boldsymbol{D}^-\boldsymbol{D}^2\boldsymbol{\beta} + \boldsymbol{e_1}(\boldsymbol{e_2} - C\boldsymbol{e_1} - C\boldsymbol{e_2})^T\boldsymbol{D}\boldsymbol{\beta}], \tag{B 11}$$

where $\boldsymbol{D}^-$ is the spectral representation of the integration operator that makes the first Chebyshev coefficient zero, and $\boldsymbol{e_k}$ is the $k^{\text{th}}$ Euclidean basis vector. (B 11) can be written in state-space form as

$$\left.\begin{aligned}\frac{\mathrm{d}}{\mathrm{d}t}\begin{bmatrix}\boldsymbol{\beta}\\\dot{\boldsymbol{\beta}}\end{bmatrix} &= \begin{bmatrix}\boldsymbol{0} & \boldsymbol{I}\\\boldsymbol{M}^{-1}\boldsymbol{A}_1 & \boldsymbol{M}^{-1}\boldsymbol{A}_2\end{bmatrix}\begin{bmatrix}\boldsymbol{\beta}\\\dot{\boldsymbol{\beta}}\end{bmatrix},\\ \boldsymbol{M} &= 2R\boldsymbol{I} + \boldsymbol{A}\boldsymbol{D}^-,\\ \boldsymbol{A}_1 &= -\frac{2}{3}S\boldsymbol{D}^4 - \boldsymbol{A}\boldsymbol{D}^-\boldsymbol{D}^2 + \boldsymbol{A}\boldsymbol{e_1}(\boldsymbol{e_2} - C\boldsymbol{e_1} - C\boldsymbol{e_2})^T\boldsymbol{D},\\ \boldsymbol{A}_2 &= -2\boldsymbol{A}\boldsymbol{D}^-\boldsymbol{D} + \boldsymbol{A}\boldsymbol{e_1}(\boldsymbol{e_2} - C\boldsymbol{e_1} - C\boldsymbol{e_2})^T.\end{aligned}\right\} \tag{B 12}$$

When numerically solving the system, the infinite series are truncated to finite series. In order to incorporate the four boundary conditions into (B 12), the last four rows of the differential equation for $\ddot{\boldsymbol{\beta}}$ are replaced by the boundary conditions. The system is then

$$\frac{\mathrm{d}}{\mathrm{d}t}\begin{bmatrix}\boldsymbol{I} & \boldsymbol{0}\\\boldsymbol{0} & \boldsymbol{I}_{-4}\end{bmatrix}\begin{bmatrix}\boldsymbol{\beta}\\\dot{\boldsymbol{\beta}}\end{bmatrix} = \begin{bmatrix}\boldsymbol{0} & \boldsymbol{I}\\\boldsymbol{M}^{-1}\boldsymbol{A}_1 & \boldsymbol{M}^{-1}\boldsymbol{A}_2\end{bmatrix}\begin{bmatrix}\boldsymbol{\beta}\\\dot{\boldsymbol{\beta}}\end{bmatrix}, \tag{B 13}$$

where $\boldsymbol{I}_{-4}$ is the identity matrix with the last four diagonal entries being zeros. The last four rows of the right hand side are replaced by the boundary conditions. We now have a generalized eigenvalue problem to solve for the eigenvalues of the system.

### B.1. *Nonlinear eigenvalue problem*

Having obtained the solution to the linear eigenvalue problem, we use it as an initial guess for the nonlinear eigenvalue problem. The nonlinear eigenvalue problem is obtained

by making the ansatz

$$\left.\begin{aligned}Y(x,t) &= e^{\lambda t}Y_0(x),\\ Y_0(x) &= \tfrac{1}{2}\beta_0 + \sum_{k=1}^{\infty}\beta_k T_k(x).\end{aligned}\right\} \quad (B\,14)$$

This is the same as in Appendix A, except that we allow the exponent $\lambda$ to be any complex number instead of just an imaginary number. Proceeding as in Appendix B, we arrive at the following equations:

$$2\lambda^2 R\boldsymbol{\beta} + \frac{2}{3}S\boldsymbol{D}^4\boldsymbol{\beta} = \boldsymbol{P}, \quad (B\,15)$$

$$\boldsymbol{P} = \boldsymbol{A}\boldsymbol{a}, \quad (B\,16)$$

$$\boldsymbol{D}\boldsymbol{a} = -\lambda^2\boldsymbol{\beta} - 2\lambda\boldsymbol{D}\boldsymbol{\beta} - \boldsymbol{D}^2\boldsymbol{\beta}, \quad (B\,17)$$

$$\boldsymbol{V} = \lambda\boldsymbol{\beta} + \boldsymbol{D}\boldsymbol{\beta}, \quad (B\,18)$$

$$a_0 = -C(\lambda)(V_0 + V_1) + V_1, \quad (B\,19)$$

where the notation is as in Appendix B.

Putting everything together, we get the following equation:

$$2\lambda^2 R\boldsymbol{\beta} + \frac{2}{3}S\boldsymbol{D}^4\boldsymbol{\beta} = \boldsymbol{A}[-\lambda^2\boldsymbol{D}^{-}\boldsymbol{\beta} - 2\lambda\boldsymbol{D}^{-}\boldsymbol{D}\boldsymbol{\beta} + \lambda\boldsymbol{e_1}(\boldsymbol{e_2} - C(\lambda)\boldsymbol{e_1} - C(\lambda)\boldsymbol{e_2})^T\boldsymbol{\beta}$$
$$-\boldsymbol{D}^{-}\boldsymbol{D}^2\boldsymbol{\beta} + \boldsymbol{e_1}(\boldsymbol{e_2} - C(\lambda)\boldsymbol{e_1} - C(\lambda)\boldsymbol{e_2})^T\boldsymbol{D}\boldsymbol{\beta}], \quad (B\,20)$$

where the notation is as in Appendix B. Truncating the upper limit of the infinite series to $N$, (B 20) gives $N+1$ equations for $N+2$ unknowns (the $N+1$ elements of $\boldsymbol{\beta}$ and $\lambda$). We add an equation which normalizes $\boldsymbol{\beta}$ in order to make the system square. As before, the last four equations are replaced by the boundary conditions. We solve for $\boldsymbol{\beta}$ and $\lambda$ using the Newton-Raphson method, using absolute and relative error tolerances $10^{-6}$. For cases where the Newton-Raphson method did not converge, we calculated the solution by looking at a global picture of the determinant of the system and finding its roots.

To validate our method for calculating eigenvalues, we calculate the eigenvalues for the same set of parameters as in figures 4c and 4d in Alben (2008*a*). In figure 21, we compare the eigenvalues calculated using our method to some of the eigenvalues in Alben (2008*a*), adopting the notation used in that work. Our eigenvalues agree well with those from Alben (2008*a*), lending confidence to our method.

### B.2. *Quiescent fluid*

Consider the case where the plate is immersed in a quiescent fluid, i.e. where the bending velocity is large compared to the fluid velocity. How do the eigenvalues of the system change? To answer this question, we solve the Euler-Bernoulli and Euler equations (2.1)–(2.2) in the limit of large bending velocity. In this limit, the appropriate time scale to use is the bending time scale, which we choose to be $\sqrt{3\rho_s L^4/(4Ed^2)}$. Nondimensionalizing the solid and fluid equations using the length scale $L/2$ and the bending time scale yields

$$\left.\begin{aligned}Y_{tt} + Y_{xxxx} &= \frac{1}{2R}\Delta p,\\ \nabla\cdot\mathbf{u} &= 0,\\ \mathbf{u}_t + \sqrt{\frac{3R}{S}}\mathbf{u}_x &= \nabla\phi,\end{aligned}\right\} \quad (B\,21)$$

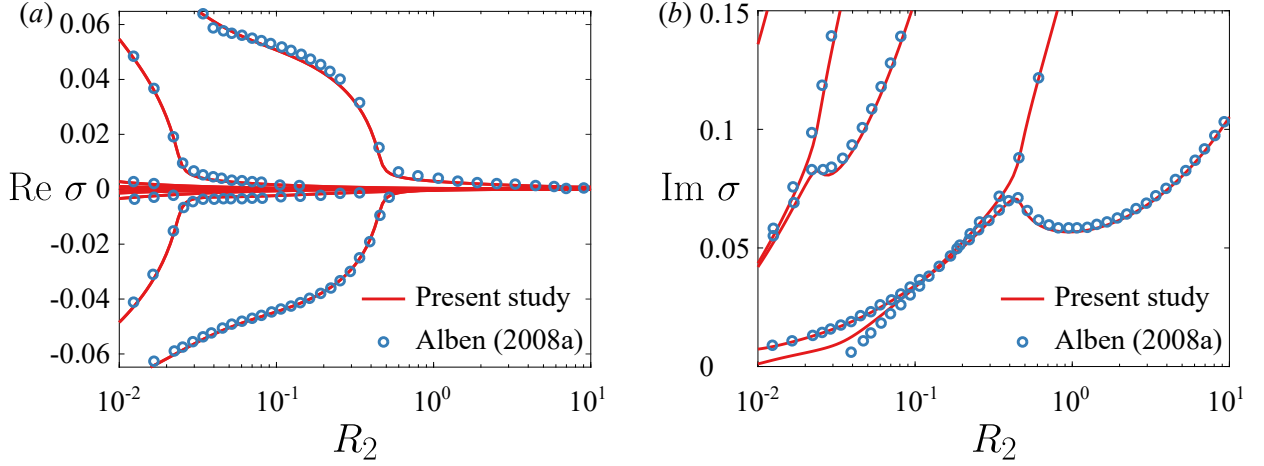

FIGURE 21. Comparison between eigenvalues calculated using our method and those found in figures 4c and 4d in Alben (2008*a*). Note that just for this figure, we adopt the notation used in Alben (2008*a*).

where $R$ and $S$ are as in (2.4), and $\phi = p_\infty - p$. In the above, $x$, $t$, $Y$, $\mathbf{u}$, and $p$ are now dimensionless, with the pressure nondimensionalized by $\rho_f E d^3/(3\rho_s d L^2)$. The limit of a quiescent flow corresponds to $R/S \to 0$, or equivalently $Ed^2/\rho_s L^2 \gg U^2$, which explicitly puts this limit in terms of velocity scales. For now, we keep all terms and discuss the limit later. Intuitively, large values of the solid-to-fluid mass ratio $R$ make the fluid dynamics inconsequential to the deflection of the plate (a heavy plate will be unaffected by the surrounding fluid).

The fluid additionally satisfies the no-penetration condition, stated as

$$v|_{x\in[-1,1],y=0} = Y_t + \sqrt{\frac{3R}{S}} Y_x. \tag{B 22}$$

The boundary conditions on the plate are

$$Y(-1,t) = 0, \quad Y_x(-1,t) = 0, \quad Y_{xx}(1,t) = 0, \quad Y_{xxx}(1,t) = 0. \tag{B 23}$$

We solve for the fluid motion for a given deflection as in Appendix A. Writing the deflection as

$$Y(x,t) = \frac{1}{2}\beta_0(t) + \sum_{k=1}^{\infty} \beta_k(t) T_k(x), \tag{B 24}$$

and the components of the complex potential evaluated on the surface of the plate as

$$\left.\begin{aligned}\phi(z = x \pm 0\mathrm{i}, t) &= \pm\frac{1}{2}a_0(t)\sqrt{\frac{1-x}{1+x}} \pm \sum_{k=1}^{\infty} a_k(t)\sin k\theta, \\ \psi(z = x \pm 0\mathrm{i}, t) &= \frac{1}{2}a_0(t) + \sum_{k=1}^{\infty} a_k(t) T_k(x),\end{aligned}\right\} \tag{B 25}$$

the pressure difference across the surface of the plate is

$$\Delta p(x,t) = a_0(t)\sqrt{\frac{1-x}{1+x}} + 2\sum_{k=1}^{\infty} a_k(t)\sin k\theta. \tag{B 26}$$

The coefficients $a_k$ are obtained by applying the no-penetration condition,

$$\frac{\partial \psi}{\partial x}\bigg|_{y=0} = -\left(\frac{\partial}{\partial t} + \sqrt{\frac{3R}{S}}\frac{\partial}{\partial x}\right)^2 Y. \tag{B 27}$$

This does not yield $a_0$, which is instead given by the Laplace domain equation

$$a_0 = -\sqrt{\frac{3R}{S}}C(V_0 + V_1) + \sqrt{\frac{3R}{S}}V_1. \tag{B 28}$$

In the limit of a quiescent fluid ($R/S \to 0$), $a_0 \to 0$. Thus all of the coefficients $a_k$ are determined by (B 27), which itself simplifies since the second term in the parentheses is zero in the limit $R/S \to 0$. We note that in this limit the only fluid force on the plate is the force due to added mass.

Putting everything together, we get the following ODE:

$$\ddot{\boldsymbol{\beta}} + \boldsymbol{D}^4\boldsymbol{\beta} = -\frac{1}{R}\boldsymbol{A}\boldsymbol{D}^{-}\ddot{\boldsymbol{\beta}}, \tag{B 29}$$

where $\boldsymbol{\beta}$ is the vector of coefficients $\beta_k$, $\boldsymbol{D}$ is the spectral representation of the differentiation operator, and $\boldsymbol{D}^{-}$ is the spectral representation of the integration operator that makes the first Chebyshev coefficient zero. The operator $\boldsymbol{A}$ maps the coefficients $a_k$, which are the coefficients of a *sine* series for the pressure, into the corresponding coefficients of a *cosine* series. If $\boldsymbol{T}_s$ is an operator which takes us from the $x$-domain to the sine domain, and $\boldsymbol{T}_c$ is an operator which takes us from the $x$-domain to the cosine domain, then $\boldsymbol{A} = \boldsymbol{T}_c\boldsymbol{T}_s^{-1}$. (B 29) can be written in state-space form as

$$\frac{\mathrm{d}}{\mathrm{d}t}\begin{bmatrix}\boldsymbol{\beta}\\\dot{\boldsymbol{\beta}}\end{bmatrix} = \begin{bmatrix}\boldsymbol{0} & \boldsymbol{I}\\-\left(\boldsymbol{I}+\frac{1}{R}\boldsymbol{A}\boldsymbol{D}^{-}\right)^{-1}\boldsymbol{D}^4 & \boldsymbol{0}\end{bmatrix}\begin{bmatrix}\boldsymbol{\beta}\\\dot{\boldsymbol{\beta}}\end{bmatrix}. \tag{B 30}$$

When numerically solving the system, the infinite series are truncated to finite series. In order to incorporate the four boundary conditions into (B 31), the last four rows of the differential equation for $\ddot{\boldsymbol{\beta}}$ are replaced by the boundary conditions. This is fine to do since the last four rows read $\ddot{\beta}_k = 0$ due to four applications of the differentiation operator $\boldsymbol{D}$. The system is then

$$\frac{\mathrm{d}}{\mathrm{d}t}\begin{bmatrix}\boldsymbol{I} & \boldsymbol{0}\\\boldsymbol{0} & \boldsymbol{I}_{-4}\end{bmatrix}\begin{bmatrix}\boldsymbol{\beta}\\\dot{\boldsymbol{\beta}}\end{bmatrix} = \begin{bmatrix}\boldsymbol{0} & \boldsymbol{I}\\-\left(\boldsymbol{I}+\frac{1}{R}\boldsymbol{A}\boldsymbol{D}^{-}\right)^{-1}\boldsymbol{D}^4 & \boldsymbol{0}\end{bmatrix}\begin{bmatrix}\boldsymbol{\beta}\\\dot{\boldsymbol{\beta}}\end{bmatrix}, \tag{B 31}$$

where $\boldsymbol{I}_{-4}$ is the identity matrix with the last four diagonals being zeros. The last four rows of the right hand side are replaced by the boundary conditions. We now have a generalized eigenvalue problem to solve for the eigenvalues of the system.

## Appendix C. Some useful formulas

The following is a collection of useful definitions and formulas from Moore (2017) for the Chebyshev method employed here. The (interior) Gauss-Chebyshev points are

$$x_n = \cos\theta_n, \quad \theta_n = \frac{\pi(2n+1)}{2(N+1)}, \quad \text{for } n = 0, 1, \ldots, N. \tag{C 1}$$

Consider a function $f(x)$ interpolated at these points by the polynomial $p_N(x)$ of degree $N$:

$$\left.\begin{aligned}f(x_n) &= p_N(x_n), \quad \text{for } n = 0, 1, \ldots, N,\\P_N(x_n) &= \frac{1}{2}b_0 + \sum_{k=1}^{N} b_k T_k(x).\end{aligned}\right\} \tag{C 2}$$

On the $\theta$-grid this is

$$f(x_n) = \frac{1}{2}b_0 + \sum_{k=1}^{N} b_k \cos k\theta_n, \quad \text{for } n = 0, 1, \ldots, N. \tag{C 3}$$

Thus we may use the fast discrete cosine transform to transform between a function's values on the collocation points, $f(x_n)$, and the Chebyshev coefficients $b_k$.

The antiderivative of $p_N(x)$ is

$$\left.\begin{aligned}D^{-1}p_N(x) &= \frac{1}{2}B_0 + \sum_{k=1}^{N+1} B_k T_k(x),\\ B_k &= \frac{1}{2k}(b_{k-1} - b_{k+1}), \quad \text{for } n = 1, 2, \ldots, N.\end{aligned}\right\} \tag{C 4}$$

$B_0$ is a free constant of integration.

The derivative of $p_N(x)$ is

$$\left.\begin{aligned}Dp_N(x) &= \frac{1}{2}b'_0 + \sum_{k=1}^{N} b'_k T_k(x),\\ b'_{N+1} &= b'_N = 0,\\ b'_k &= b'_{k+2} + 2(k+1)b_{k+1}, \quad \text{for } n = N-1, N-2, \ldots, 0.\end{aligned}\right\} \tag{C 5}$$

Since the endpoints $x = \pm 1$ are not part of the collocation grid, we give a formula to evaluate the function at the endpoints:

$$p_N(\pm 1) = \frac{1}{2}b_0 + \sum_{k=1}^{N} (\pm 1)^k b_k. \tag{C 6}$$